\documentclass[preprint,nopreprintline,11pt]{elsarticle}

\usepackage{amssymb}
\usepackage{amsthm}
\usepackage{amsmath}
\usepackage{lineno}
\usepackage[utf8]{inputenc}
\usepackage[T1]{fontenc}
\usepackage{xcolor, soul}
\usepackage{url}
\usepackage[margin=3cm]{geometry}
\usepackage{subcaption}
\usepackage{siunitx}
\usepackage{booktabs}
\usepackage{tabularx}
\usepackage{wrapfig}
\usepackage{mathtools}
\usepackage{wrapfig}

\journal{Artificial Intelligence in Medicine }
\colorlet{lred}{red!40}
\colorlet{lgreen}{green!40}
\colorlet{lblue}{blue!40}
\definecolor{bananamania}{rgb}{0.98, 0.91, 0.71}
\newcommand{\mc}[1]{\mathcal{#1}}
\newcommand{\R}{\mathbb{R}}
\newcommand{\norm}[1]{\left\lVert#1\right\rVert}

\newcommand{\fidinc}{$\text{FID}_\text{inc}$}
\newcommand{\fidres}{$\text{FID}_\text{res}$}
\newcommand{\fidvgsa}{$\text{FID}_\text{vgs}$}
\newcommand{\rinc}{$\mathbf{r_\text{inc}}$}
\newcommand{\rres}{$\mathbf{r_\text{res}}$}
\newcommand{\rvgsa}{$\mathbf{r_\text{vgs}}$}

\DeclareMathOperator*{\argmin}{arg\,min}
\DeclareMathOperator*{\argmax}{arg\,max}

\begin{document}
	
	\begin{frontmatter}
		
		
		
		\title{Three-dimensional Bone Image Synthesis with Generative Adversarial Networks}
		
		\affiliation[label1]{
			organization={VASCage - Centre on Clinical Stroke Research},
			addressline={Adamgasse 23},
			city={Innsbruck},
			postcode={6020},
			country={Austria}}

		\affiliation[label2]{organization={Core Facility Micro-CT, University Hospital for Radiology},
			addressline={Anichstra{\ss}e 35},
			city={Innsbruck},
			postcode={6020},
			country={Austria}}
		
		\affiliation[label3]{
			organization = {University Hospital for Orthopedics and Traumatology},
			addressline={Anichstra{\ss}e 35},
			city={Innsbruck},
			postcode={6020},
			country={Austria}}
		
		\affiliation[label4]{organization={Department of Mathematics, Universit{\"a}t Innsbruck},
			addressline={Technikerstra{\ss}e 13},
			city={Innsbruck},
			postcode={6020},
			country={Austria}}

		\author[label1]{Christoph Angermann}
		\author[label1,label2]{Johannes Bereiter-Payr}
		\author[label3]{Kerstin Stock}
		\author[label4]{Markus Haltmeier}
		\author[label2]{Gerald Degenhart}

		%
		
		\begin{abstract}
			Medical image processing has been highlighted as an area where deep learning-based models have the greatest potential. However, in the medical field in particular, problems of data availability and privacy are hampering research progress and thus rapid implementation in clinical routine. The generation of synthetic data not only ensures privacy, but also allows to \textit{draw} new patients with specific characteristics, enabling the development of data-driven models on a much larger scale. This work demonstrates that three-dimensional generative adversarial networks (GANs) can be efficiently trained to generate high-resolution medical volumes with finely detailed voxel-based architectures. In addition, GAN inversion is successfully implemented for the three-dimensional setting and used for extensive research on model interpretability and applications such as image morphing, attribute editing and style mixing. 
			The results are comprehensively validated on a database of three-dimensional HR-pQCT instances representing the bone micro-architecture of the distal radius.
			
		\end{abstract}
		
		
		
		
	\end{frontmatter}

	\section{Introduction}
	\label{sec:introduction}
	
	The adoption of Deep Learning (DL) into the broad field of medical imaging is an ongoing and remarkable success story. From decision support systems in radiology \cite{sahiner2019}, over segmentation algorithms for complex organ and tumour regions \cite{gruber2022,lenchik2019} to applications for image enhancement and super-resolution \cite{mahapatra2019}, the use of learning-based techniques has led to many advances with great potential for future applications. Such applications require the availability of large amounts of training data to ensure a sufficient range of population variability and thus  to increase the reliability of the developed models \cite{fetty2020}. When it comes to development of medical applications, the availability of sufficient data in the relevant modalities is often limited. In addition, sharing medical data with other institutions or even between different hospitals is a major challenge for legal and privacy reasons \cite{ching2018}. These limitations make it challenging to integrate existing modern methods into routine clinical practice.

	\subsection{Generative modeling}
	
	A promising approach to overcome above mentioned challenges is the synthetic generation of realistic targeted data samples. This not only ensures patient privacy, but also allows new types of images with specific characteristics to be synthesised on demand, enabling medical research on a much larger scale. Within the field of generative modelling, the advent of generative adversarial networks (GANs) in 2014 can be seen as a major catalyst \cite{goodfellow2014,wang2020}. GANs have significantly advanced a wide range of life science applications \cite{fetty2020,burlina2019} as well as other areas within medical imaging, including modality transfer \cite{angermann2023, wolterink2017} and image segmentation \cite{pelaezvegas2023}. Generative models approximate the probability density function underlying the available data and can thus produce realistic representations of examples that differ from those in the training data \cite{pinaya2022}. 
	GANs have achieved remarkable improvements in the quality of natural images \cite{karras2017,karras2020}, and also allow for good control of output diversity and resolution. In addition, the introduction of GAN inversion techniques has allowed a variety of new possibilities beyond synthesis, such as attribute manipulation, image transitions, and style mixing, to name a few \cite{xia2022}.\\

	A major challenge in using generative models for medical applications is the dimensionality of the data. Existing GANs are mainly built and tested on large data sets of two-dimensional images, such as the CelebA-HQ data set (30k face portraits) \cite{karras2017} or LSUN (10 scene and 20 object categories with at least 125k images in each category) \cite{yu2015}. Key research in medical imaging, however, is often carried out on three-dimensional data (3D volumes). Compared to two-dimensional data (2D images), this allows a more precise interpretation of the objects of interest by exploiting their 3D structure and information. The number of voxels is typically much higher than the number of pixels in the two-dimensional counterparts, and processing 3D networks becomes a major challenge. In addition, the lack of large amounts of patient data further limits the applicability of state-of-the-art generative 2D models to the 3D case.

	\subsection{Case example: 3D Bone Image Synthesis}
	
	An example highlighting the need for 3D generative models is the analysis of bone micro-architecture structure.  High-resolution peripheral quantitative computed tomography (HR-pQCT) is a 3D medical imaging technique capable of examining in vivo microscopic bone structures in the extremities. Since its introduction in 2005 \cite{boutroy2005}, its use in clinical research into bone-related pathologies has grown rapidly due to the unprecedented resolution of the images \cite{whittier2020}. With \SIrange{3}{5}{\micro\sievert} effective radiation dose per scan, HR-pQCT is also beneficial to patients compared to conventional (diagnostic) bone imaging techniques such as dual-energy X-ray absorptiometry (DXA), while providing significantly more valuable information about overall bone quality \cite{whittier2023}.\\

	\begin{figure}[htb!]
		\centering
		\begin{subfigure}{.1\textwidth}
			\includegraphics[width=.99\textwidth]{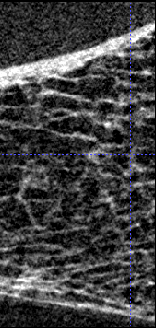}\vspace{5.5em}
		\end{subfigure}\hspace{.1em}
		\begin{subfigure}{.315\textwidth}
			\includegraphics[width=.99\textwidth]{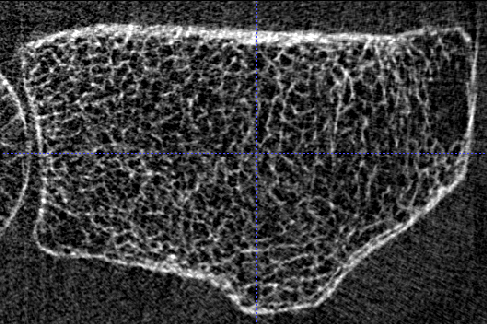}\\[.15cm]
			\includegraphics[width=.99\textwidth]{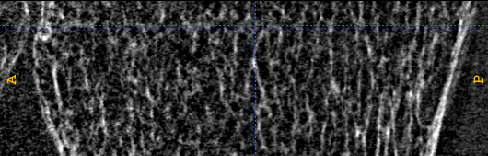}\vspace{.9em}
		\end{subfigure}\hspace{4em}
		\begin{subfigure}{.118\textwidth}
			\includegraphics[width=.99\textwidth]{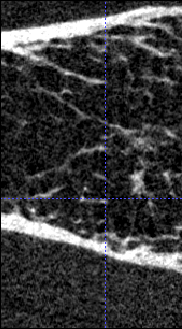}\vspace{5.25em}
		\end{subfigure}\hspace{.1em}
		\begin{subfigure}{.317\textwidth}
			\includegraphics[width=.99\textwidth]{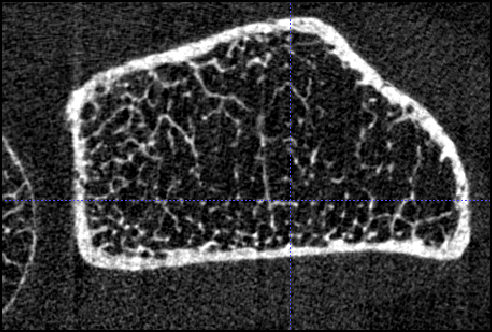}\\[.15cm]
			\includegraphics[width=.99\textwidth]{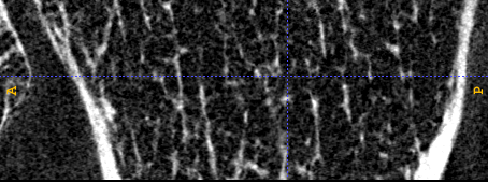}
		\end{subfigure}
		\caption{HR-pQCT bone samples of real patients with isotropic voxel size \SI{60.7}{\micro\metre}. Volumes are cropped to a region of interest (ROI) with varying number of voxels for each scan.}
	\end{figure}
	
	Despite the clear advantages, the current use of HR-pQCT still remains confined to research applications. Major obstacles to its adoption into clinical diagnostic routine are the time-consuming segmentation process \cite{buie2007, neeteson2023} and the large number of interdependent parameters generated by bone morphometric analysis \cite{whittier2023}. Both issues have been addressed by the use of machine learning as detailed in recent publications \cite{whittier2023, neeteson2023}. However, to our knowledge, all existing large cohorts of patient data have been recruited for the study of bone-related pathologies (see \cite{samelson2019} as an example). This limits researchers attempting to train and verify their models with HR-pQCT volumes of bones from young, non-pathological patients to small data sets consisting of structures from only a few individuals.

	\subsection{Main contributions}
	
	The results of this work provide a pathway to overcome such limitations by generating arbitrary amounts of 3D volumes with a tailored set of relevant properties. It bridges the gap between recent advances in two-dimensional generative modelling and their implementation for high-resolution 3D medical volumes. To this end, the techniques of progressive growing (ProGAN) \cite{karras2017} and style-based generation (StyleGAN) \cite{karras2020} are extended to the 3D case. The GAN model and the entire training algorithm are developed from scratch in PyTorch (\url{https://pytorch.org/}). 
	\\
	
	Our approach is implemented on a modest sample set of 404 bone volumes obtained through HR-pQCT. The result is a powerful high-resolution bone image synthesis model of surprisingly good quality and diversity. Specifically, 64 synthetic instances are assessed by two CT imaging experts. In addition, advanced visual assessment metrics taken from computer vision are implemented and compared to the expert assessment of the generated bone images. This can be used as an expert-driven indicator of how a computer best mirrors human visual perception.\\
	
	To gain a more detailed understanding of the structure of the 3D model, the latent codes of the model are examined in further detail. Specific attributes of the data and the corresponding latent inputs are used to learn directions in latent space that describe these attributes well. In addition, GAN inversion techniques and latent code manipulation are explored to synthesise customised high-dimensional medical images for attribute-driven data augmentation. The results are supported by a large number of visualisations in the results section, which also includes links to demonstration videos of the proposed analysis of 3D generative models. To ensure reproducibility, exact details on the optimisation process and the network architectures are summarised in the supplementary material. An extensive literature search revealed that this is the first work on generating highly detailed bone micro-architecture in 3D. Furthermore, this is the only work to date that investigates latent space properties and automated realism assessment in 3D medical applications.

	\section{Background}
	\label{sec:related_work}
	
	\subsection{Generative Adversarial Networks}
	In basic terms, a generative adversarial model learns a link function between a low-dimensional latent distribution and a high-dimensional data distribution. The GAN architecture \cite{goodfellow2014} is composed of a generator function $G:\mc Z \to \mc X$ and an adversarial counterpart $f: \mc X \to [0,1]$. The elements of latent space $\mc Z$ are commonly assumed to follow a standard normal distribution, i.e., the generator takes a sample $z\in \mc Z,\ z\sim \mc N(0,1)$ and maps it to image space $\mc X$. The generative function $G$ is approximated by a neural network, by adaptation of its parameters so that the output distribution of $G$ assimilates the distribution of the given training set. Simultaneously, the adversarial function $f$ is optimized to distinguish between generated and real instances. In a two-player min-max game, generator parameters are updated to fool a steadily improving discriminator \cite{angermann2023}. Already the initial versions of GANs raised significant interest in the computer vision community, but proved to be unstable due to the problem of the vanishing gradients and mode collapse. Improving the optimization objective of the generative and adversarial function yielded highly successful modifications of the simple two-player game, like Least-Squares-GAN \cite{mao2017}, Spectral-Normalization-GAN \cite{miyato2018} or Wasserstein-GAN (WGAN) \cite{gulrajani2017,arjovsky2017}. Especially the WGAN approach had a crucial impact on training controllability and substantially shaped GAN development. Instead of classifying if a sample is real ($f\approx 1$) or has been drawn by a neural network ($f\approx 0$), Wasserstein GANs use a new adversarial critic  $f:\mc X \to \R$ to approximate the distance between the real and the generator distribution.
	
	\subsection{High-Resolution Synthesis}
	The desire to draw synthetic images in higher resolutions led to introduction of progressive GAN (ProGAN) \cite{karras2017}, which uses a growing strategy for the network training process. The core concept is to start with low resolution for both generative and adversarial functions and then add new layers as training progresses, modelling fine high-frequency details \cite{xia2022}. ProGAN improved both the optimization speed and the stability, facilitating image generation at a resolution of $1024^2$ pixels. Controlling the style of synthetic images became increasingly important and has been successfully assessed by style-based GAN (StyleGAN) \cite{karras2019}. The model manipulates mean and variance per channel after each convolution in the generative function to control the style of the output effectively and, similar to ProGAN, enables generation up to a scale of $1024^2$ pixels. Improvement of perceptual quality was achieved in StyleGAN2 \cite{karras2020} by including weight demodulation, path length regularization and network architecture redesign. Embedding adaptive discriminator augmentation in StyleGAN2-Ada \cite{karras2020ada} yielded reasonable training of style-based generators also on limited data sets. Latest progress has been made in StyleGAN3 \cite{karras2021} that proposed a new architecture to tackle aliasing effects during image transition.
	
	\subsection{GAN Inversion}
	ProGAN and StyleGAN enable a meaningful link between image space and a corresponding latent and style vectors, respectively. Beside the unconditioned generation of images, these models may also be used for semantic manipulation and effective augmentation of existing data. GAN inversion aims to invert a given instance from data space back into its latent or style representation, so that the image can be reconstructed from the inverted code by the pretrained generative function. GAN inversion plays a critical role in bridging the real and synthetic data domains, leading to significant advances in this fairly young research area \cite{xia2022, tov2021, shen2021,zhu2020domain}. So far, the rapidly-growing set of solutions for GAN inversion has been divided into three sub-areas. 
	
	Learning-based inversion is characterized by the use of an additional encoding neural network which predicts the latent code from an existing image such that the GAN-based reconstructed counterpart resembles the original. Optimization-based methods directly minimize a pixel-wise reconstruction loss to find a corresponding latent code for an existing image. The minimization objective is commonly solved by gradient descent method. Both techniques lead to a quality-to-time trade-off \cite{xia2022} -- learning-based methods are generally associated with quality degradation of the reconstruction, while optimization-based methods are time-consuming and strongly depend on the initial value for the minimization algorithm. Therefore, hybrid methods are the most widely adopted methods to date, using an encoder-based latent code as the starting value for the subsequent optimization process.

	\subsection{GANs in Medical Imaging}
	GAN synthesis and inversion has already been adopted by the medical community, where existing methods for inversion and manipulation are used in specific domains like computed tomography (CT) or magnetic resonance imaging (MRI).  In \cite{ren2021}, the idea of domain-specific GAN inversion \cite{zhu2020domain} is incorporated to synthesize mammograms constrained on shape and texture for psychophysical analysis on larger scale. In \cite{fetty2020} a StyleGAN is trained on both, CT and MRI instances, and it is shown how specific attributes can be targeted in the latent space, enabling powerful methods for guided manipulation and modality transfer. While the previously mentioned works only use 2D slices, \cite{hong2021} targets entire stacks of images, using a 3D-StyleGAN to synthesize MRI images. Although this work demonstrates StyleGAN adoption to 3D data, the authors limit data dimension to $64^3$ voxels -- a size that is rarely sufficient in real-life medical studies. Furthermore, no analysis on latent code interpretation and manipulation has been made. 
	
	Most closely related to our study is the hierarchical amortized GAN (HA-GAN) proposed in \cite{sun2022}. A hierarchical structure is implemented that simultaneously generates a low-resolution version of the 3D dataset and a randomly selected sub-volume of the high-resolution counterpart. In terms of 3D synthesis at high resolution, this work achieves tremendous performance. However, the semantic meanings of the latent space are explored exemplary by implementing two additional regression problems. Furthermore, the authors of HA-GAN do not emphasise advanced feature extraction to model the realism of the generated samples. These aspects clearly distinguish HA-GAN from the study presented here.

	\section{Methods}
	\label{sec:methods}
	
	In the present work, two methods for volumetric synthesis are considered: 3D progressive growing GAN (3D-ProGAN) and 3D style-based GAN (3D-StyleGAN). These methods are described in section \ref{ssec:arch} and applied
	to a data set described in section \ref{ssec:data}. A short glance on the used visual validation metrics is given section \ref{ssec:vali}. Section \ref{ssec:training} includes some important details on model training and section \ref{ssec:inversion} describes the GAN inversion process.
	
	\subsection{Data Acquisition and Preprocessing}
	\label{ssec:data}
	The data set used for experimentation was obtained from a study on bone health and fracture healing conducted by the Medical University of Innsbruck in collaboration with the department for trauma surgery at the University Hospital of Innsbruck. Subjects were recruited from patients admitted to the emergency outpatient unit due to a fracture of the distal radius. In the course of the study, the fractured and non-fractured radii were scanned at six time points within one year. The non-fractured radii were scanned according to a fixed distance protocol (see \cite{whittier2020} for details), approximately \SI{10}{\milli\meter} from the distal end of the bone. The intervals were at date of admission as well as after one week, three weeks, three months, six months and 12 months -- resulting in six distinct volumes per patient. Only volumes from the non-fractured site were used from 98 patients, 515 3D volumes for the data set in total. A fifth of the volumes was removed before training due to issues with scanning quality, reducing the volume count further to 404 (cf. section \ref{ssec:vali}).\\
	
	\begin{wrapfigure}{l}{.5\textwidth}
		\centering
		\includegraphics[width=.1\textwidth]{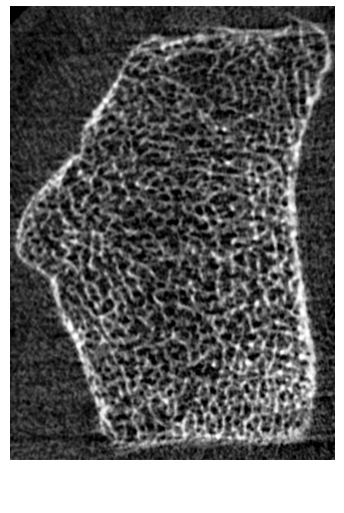}\hspace{-.4em}
		\includegraphics[width=.13\textwidth]{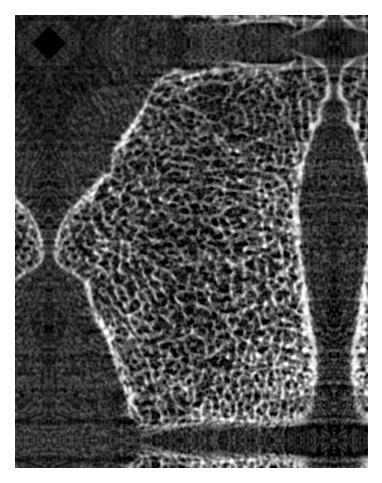}\hspace{-1em}
		\includegraphics[width=.13\textwidth]{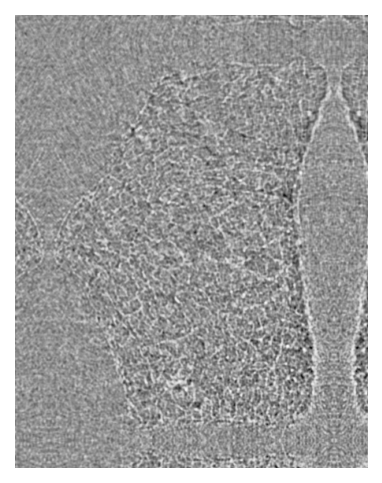}\hspace{-1em}
		\includegraphics[width=.13\textwidth]{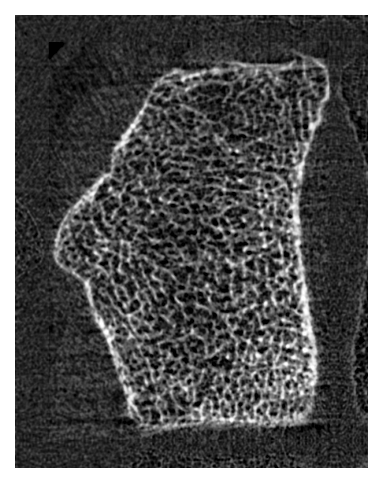}
		\caption{Preprocessing. From left to right: The sample is cropped or padded to a constant size of \numproduct{168 x 576 x 448} voxels. The mirrored volume is used as padding. The samples are considered with regard to the discrete cosine basis. Clipping the basis coefficients to range $[-1000,1000]$ yields the noise volume. The padded regions are replaced by the corresponding noise volume.}
		\label{fig:prepro}
	\end{wrapfigure}
	
	A strong variation of measured voxels between the individual measurements makes the data processing a non-trivial task. While 168 axial slices ($\approx$ \SI{10}{\milli\meter}) were obtained for every sample, the extent in vertical and horizontal direction ranges between $[397, 663]$ and $[278,529]$ voxels, respectively. 
	The processing pipeline consists of multiple steps and is shown in Figure \ref{fig:prepro}. Each sample is cropped or padded to a constant size of \numproduct{168 x 576 x 448} voxels. The mirrored image is used as padding, as conventional zero padding is not appropriate in this case due to the high levels of background noise. The samples are considered with regard to the discrete cosine basis. Clipping the basis coefficients to range $[-1000,1000]$ yields the noise images. The padded regions are replaced by the corresponding noise image to avoid reflections of the bone itself at the edges. Due to restricted hardware resources, patient data is sub-sampled by factor 2. \\
	
	Following the described pre-processing pipeline, training data is transformed to a unique shape of \numproduct{84x288x224} with constant voxel spacing. To further enlarge the dataset, each scan is divided into four overlapping slice stacks of size \numproduct{32x288x224}. This is followed by rotations and zoom-in operations using angles in $[-10,10]$ and zoom factors in $[1,1.15]$, both uniformly chosen at random. Using the augmentation pipeline described above, nearly 6800 training instances are obtained from the 404 volumes considered.

	
	\subsection{Architecture}
	\label{ssec:arch}
	
\begin{figure}[thb!]
		\centering
		\includegraphics[width=.9\textwidth]{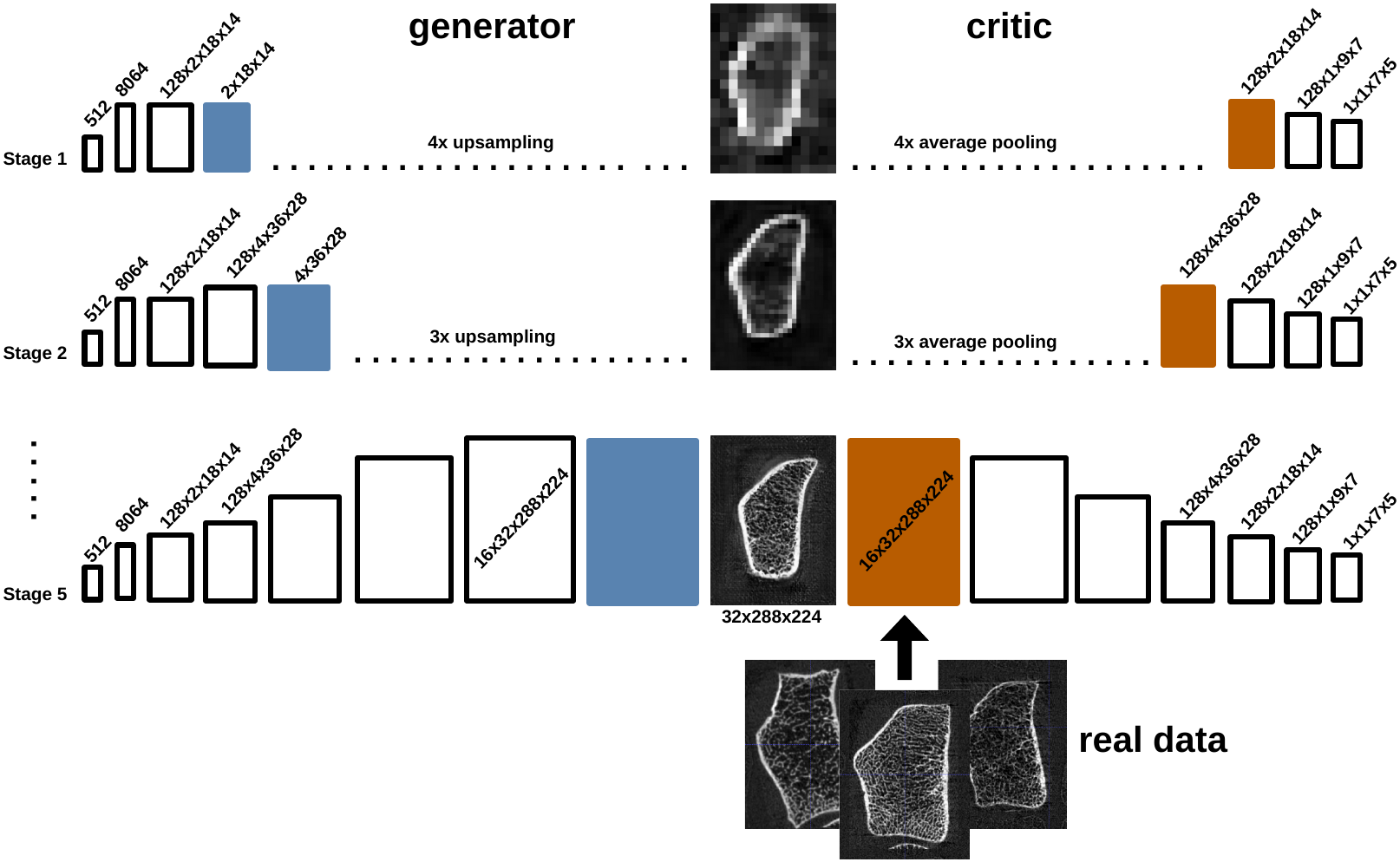}
		\caption{Exemplary visualization of the progressive growing strategy for synthesis of 3D bone HR-pQCT data.}
		\label{fig:3dprogan}
		
	\end{figure}

	\textbf{3D Progressive Growing GAN:}\\
	The generator $G: \mc Z\to \mc{X}$ maps from latent space to image space. To be more precise, a normally distributed latent vector $z \in \mc Z\subset\mathbb{R}^{512},\ z \sim \mathcal{N}(\vec{0}, \text{Id})$ is sampled and forwarded to a dense layer and a reshape layer with output size $[c\cdot 8,d_1/32,d_2/32,d_3/32]$, where $c$ denotes the channel size of the method and $d_1, d_2, d_3$ the spatial size of the training data. This is followed by nearest-neighbor upsampling and a block of consecutive 3D convolutional layers. The generator is now called to reside on stage 1. A repeated application of the same block (upsampling and convolutional block) yields the stage 2 output. In total the block is applied 5 times, yielding a final output resolution of $[c, d_1, d_2, d_3]$ at stage 5 (see Figure \ref{fig:3dprogan}). In stages 3 to 5 the feature maps are decreased by factor 2, yielding channel size $c$ at the last stage. Layers shown in blue denote 3D convolution with channel size 1 to transfer learned features to the image domain.
	
	The smooth transition strategy of \cite{karras2017} is applied. As shown in Figure \ref{fig:3dprogan}, the critic also operates in different stage modes, where the final critic at stage 5 consists of five strided convolutional layers with increasing channel size and a final output convolutional layer of channel size 1 (cf. PatchGAN \cite{isola2017}). Layers shown in orange denote 3D convolution with channel size $c$ to link the image domain with the feature space.\bigskip
	
	\textbf{3D Style-based GAN:}\\
	For style-based generation, the generative function can be described by the composition  $G=\tilde G \circ \Phi: \mc Z \to \mc X$.
	Similar to 3D-ProGAN, a normally distributed vector $z\in \mc Z\subset\R^{512}$ is sampled and then mapped by a mapping network $\Phi: \mc N(\vec{0},\text{Id}) \to \mc W$ to a learned intermediate latent space $\mc W \subset \R^{512}$ which more faithfully reflects the training data distribution compared to standard normal distribution \cite{tov2021}. The latent code $w=\Phi(z)$ is converted to 15 different style codes by learned affine transformations. Incorporating the progressive GAN described previously, these 15 style vectors are fed to the generator $\tilde G:\mc{W}\to \mc X$ using weight demodulation \cite{karras2019}, three styles at each stage. After each convolution layer, a noise map is sampled of same spatial size, scaled by a single learnable parameter and added to each feature map.  The critic network for the style-based generator remains unchanged compared to 3D-ProGAN.\\
	
	For both methods, a video-demonstration of the progressive growing strategy can be viewed online:
	\begin{itemize}
		\item \url{https://www.youtube.com/watch?v=Dicd6cEaZp8} (3D-ProGAN)
		\item \url{https://www.youtube.com/watch?v=TbKN0CPWvHE} (3D-StyleGAN)
	\end{itemize}

	\subsection{Validation}
	\label{ssec:vali}
	
	In order to quantitatively evaluate perceptual quality of intermediate training samples and final results, Frech\'et Inception Distance (FID) \cite{heusel2017} is measured between the distributions of real and synthesized data. 
	FID relies on features extracted from original and synthesized instances, where the feature extractor plays an essential role and should be chosen appropriately for the task. This study considers three feature extractors:
	\begin{enumerate}
		\item The originally proposed FID  relies on the Inception v3 classification network that was pre-trained on 2D images from ImageNet \cite{deng2009}, so this measure is not directly applicable to 3D data. Therefore, from each scan, two axial slices at random positions are selected and used for FID validation. This measure is denoted by \textbf{\fidinc}.
		
		\item Similar to HA-GAN \cite{sun2022}, a 3D ResNet model pre-trained on 3D medical images \cite{chen2019} is deployed to collect features of the 3D volumes directly. This version is denoted by \textbf{\fidres}.
		
		\item Each scan of the 98 patients was evaluated directly after measurement by a medical expert for motion artefacts and given a visual grading score (VGS) score between 1 (best) and 5 (worst), as described by Sode et al. \cite{sode2011} and reiterated by Whittier et al. \cite{whittier2020}. Using this rating, a 3D ResNet classifier has been trained. FID using features by the VGS classifier are denoted with \textbf{\fidvgsa}. Images with a score of 4 or 5 were excluded from GAN training, to avoid the network replicating motion artefacts. 
	\end{enumerate}\vspace{1em}
	
	The FID has been shown to reflect human opinion of perceptual quality quite well. However, the FID may also increase when the perceptual quality is sufficiently good but the synthesis variance is decreasing. Therefore, two additional indicators for synthesis quality are added --  precision and recall \cite{kynkaanniemi2019}. Precision quantifies the percentage of generated images that are similar to training data (sufficient perceptual quality) while recall models the percentage of training data that can be recreated by the generator (coverage of the real data distribution). For precision and recall evaluation, only features extracted by the 3D medical ResNet  model are considered.\\
	
	FID, precision and recall scores compare the distributions of two data sets. Thousands of instances are sampled from both distributions and corresponding features are used to calculate the scores. Since these are quantitative measures, assessing the plausibility of a single generated sample automatically is not possible and requires human intervention. To evaluate the proposed bone synthesis with regard to the measure of realism for single instances, a realism score \cite{kynkaanniemi2019} is adopted. More precisely, the degree of realism increases the closer the features of a generated sample are to the manifold formed by the features of the real training data, and decreases otherwise.
	Similar to FID calculation, three different methods are considered for feature extraction, yielding three different realism scores: $\mathbf{r_\text{inc}}$, $\mathbf{r_\text{res}}$ and $\mathbf{r_\text{vgs}}$. All three feature extraction methods are compared with the subjective assessment of two human experts on HR-pQCT imaging to determine the realism score that most closely matches human perception.

	\subsection{Training}
	\label{ssec:training}
	Similar to \cite{karras2017}, Wasserstein loss with a two-sided gradient penalty \cite{gulrajani2017} is deployed to train both the generator and the critic in parallel. Let $P_\mc{X}$ denote the data distribution of bone images, $G$ a generator in $\{\text{3D-ProGAN, 3D-StyleGAN}\}$ and $f:\mc{X}\to\R$ the corresponding critic. Then
	
	\begin{align}
		\label{eq:critic_loss}
		\ell_\text{critic} &= \mathbb{E}_{\substack{x\sim P_\mc{X}\quad\,\,\,\\z\sim\mc N (0,\text{Id})}}\left[ f(G(z))-f(x)+p_1\cdot\left(\left(\norm{\nabla_{\tilde{x}} f(\tilde x)}_2-1\right) \right)^2 + p_2\cdot f(x)^2
		\right]\\
		\label{eq:gen_loss}
		\ell_\text{generator} &= \mathbb{E}_{z\sim\mc N (0,\text{Id})}\left[-f\left(G(z)\right)\right],
	\end{align}
	
	where $p_1$ and $p_2$ denote the influence of the gradient and drift penalty, respectively. Note that $\tilde{x}$ denotes arbitrary transition between real and generated domain \cite{gulrajani2017}. The Adam optimizer is used to minimize both objectives in \eqref{eq:critic_loss} and \eqref{eq:gen_loss}. Optimal architecture and optimizer configurations can be found in \ref{app:arch} and \ref{app:train}, respectively. Approximately \SI{10}{\percent} from the available training data in \ref{ssec:data} was excluded for early detection of critic overfitting during the training process.

	\subsection{GAN Inversion}
	\label{ssec:inversion}
	In order to investigate properties and directions in the latent space, an encoder is built to generate latent codes from existing images, i.e., inverting the generator that has been trained in 3D-ProGAN and 3D-StyleGAN. The encoder has the reversed structure of the generator (cf. Table \ref{tab:3dprogan}). Pixel feature normalization is removed and for 3D-StyleGAN inversion, two fully connected layers with leaky ReLU activation are added at the bottom of the encoder. Using a pre-trained generator $G$ and the corresponding adversarial critic $f$, the optimization of the encoder $E:\mc{X} \to \R^{512}$ closely follows \cite{tov2021}.\\
	
	For 3D-ProGAN a hybrid approach is used, i.e., an initial guess for the latent code is obtained by propagation through the learned encoder while refinement of the given code is enabled by a subsequent minimization task. Let $f_{L-1}$ denote the penultimate convolution layer of the adversarial critic $f$. Three loss terms for distortion (\textit{dist}), perceptual similarity (\textit{perc}) and latent code plausibility (\textit{latent}) are defined as follows:
	
	\begin{align}
		&\ell_\text{dist}(x,E) = \frac{0.5}{\text{\small\#voxel}}\sum_p^{\text{\#voxel}}\left(x_p - G\left(E(x)\right)_p\right)^2, \\\
		&\ell_\text{perc}(x,E) = \frac{0.5}{\text{\small\#features}}\sum_q^\text{\#features}\left(f_{L-1}(x)_q-f_{L-1}\left(G(E(x))\right)_q\right)^2,  \\\
		&\ell_\text{latent}(x,E) = \frac{1}{1024}\sum_{r=1}^{512}\left(E(x)_r\right)^2.
	\end{align}
	
	The risk functional for the encoder $E$ and the optimization objective that yields the optimal latent code $z_\text{opt}(\hat x)$ for a given image $\hat x\in\mc{X}$ are defined as:
	
	\begin{align}
		\label{eq:enc_loss}
		\ell_\text{encoder} &= \mathbb{E}_{x\sim P_\mc{X}}\bigg[\ell_\text{dist}(x,E) + \ell_\text{perc}(x,E) + \ell_\text{latent}(x,E) \bigg], \\\
		\label{eq:min}
		z_\text{opt}(\hat x) &= \argmin_{\substack{z\in \R^{512}}}\bigg[
		\frac{1}{\text{\small\#vox}}\norm{\hat x - G\left(z\right)}_2^2 +
		\frac{1}{\text{\small\#feat}}\norm{f_{L-1}(\hat x)-f_{L-1}\left(G(z)\right)}_2^2  +
		\frac{1}{512}\norm{z}_2^2\bigg].
	\end{align}
	
	During encoder training, the loss functional in \eqref{eq:enc_loss} is minimized using Adam algorithm with hyper-parameters $(\alpha,\beta_1,\beta_2)=(\num{3e-3},0.5, 0.9)$. For the optimization in \eqref{eq:min}, the Adam algorithm is again used for 100 updates with a learning rate equal to \num{7e-3}.\\
	
	For 3D-StyleGAN a similar hybrid approach is considered with a modified functional for latent code plausibility. For style-based generation, the latent codes are not assumed to follow a multivariate normal distribution but the sampled vectors are mapped to a learned latent space $\mc{W}$ by the mapping $\Phi: \mc{N}(\vec{0},\text{Id})\to \mc{W}$ and then forwarded to image space by generator $\tilde G:\mc W \to \mc X$. Therefore, given a real image, the retrieved latent code should also reside in the learned latent space. Analogous to \cite{tov2021}, a latent discriminator $D_\mc{W}:\mathbb{R}^{512}\to[0,1]$ is trained to distinguish between latent codes constructed by the encoder (fake codes) and by the mapping $\Phi$ (real codes). The loss functional for latent code plausibility is adapted as follows:
	
	\begin{equation}
		\label{eq:lat_disc}
		\ell_\mc{W}(x,E) = -\frac{1}{512}\sum_{r=1}^{512}\log\left(D_\mc{W}\left(E(x)\right)\right).
	\end{equation}
	
	In the case of style-based bone synthesis, the risk functional for the encoder $E$ and the optimization objective that yields the optimal latent code $z_\text{opt}(\hat x)$ are defined as:
	
	\begin{align}
		\label{eq:enc_loss_style}
		\ell_\text{encoder} &= \mathbb{E}_{x\sim P_\mc{X}}\bigg[5\cdot \ell_\text{dist}(x,E) + \ell_\text{perc}(x,E) + 0.04\cdot\ell_\mc{W}(x,E) \bigg], \\\
		\label{eq:min_style}
		w_\text{opt}(\hat x) &= \argmin_{\substack{w\in \R^{512}}}\bigg[
		\frac{1}{\text{\small\#feat}}\norm{f_{L-1}(\hat x)-f_{L-1}\left(\tilde G(w)\right)}_2^2 \bigg].
	\end{align}
	
	Technical details and parameters for \eqref{eq:enc_loss_style} and \eqref{eq:min_style} are the same as for 3D-ProGAN.

	\section{Results \& Discussion}
	\subsection{Image Quality}
\begin{figure}[thb!]        
		\includegraphics[height=.09\textheight]{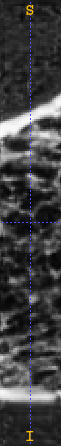}\hspace{-.1em}
		\includegraphics[height=.09\textheight]{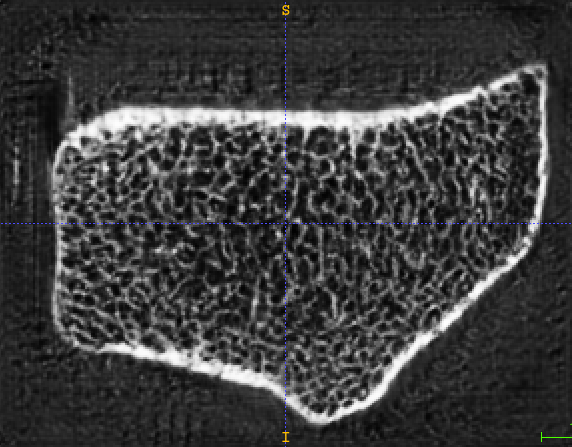}\hspace{.5em}
		\includegraphics[height=.09\textheight]{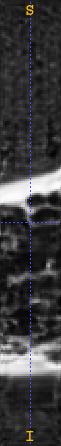}\hspace{-.1em}
		\includegraphics[height=.09\textheight]{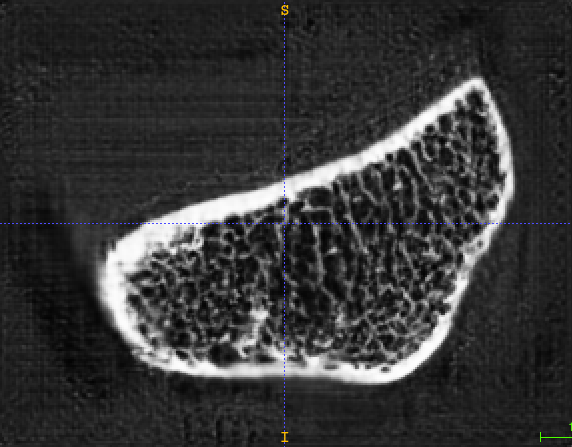}\hspace{.5em}
		\includegraphics[height=.09\textheight]{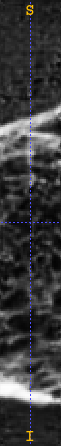}\hspace{-.1em}
		\includegraphics[height=.09\textheight]{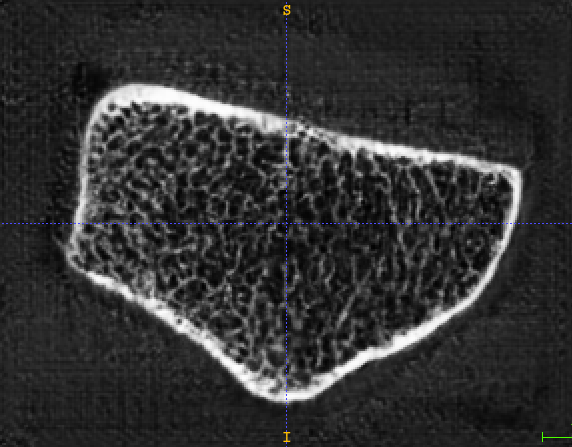}\hspace{.5em}
		\includegraphics[height=.09\textheight]{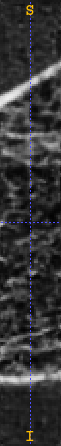}\hspace{-.1em}
		\includegraphics[height=.09\textheight]{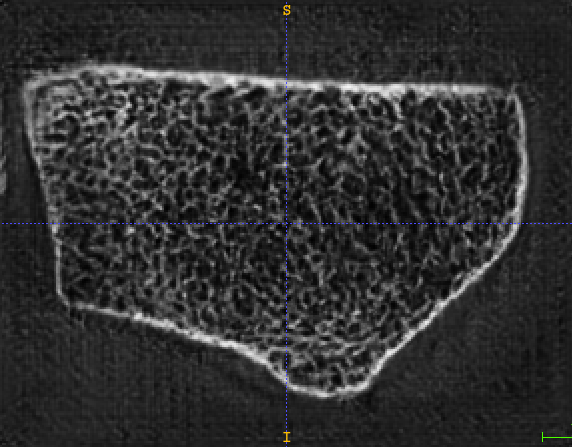}\hspace{.5em}
		\includegraphics[height=.09\textheight]{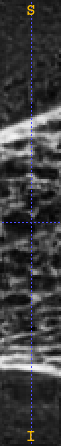}\hspace{-.1em}
		\includegraphics[height=.09\textheight]{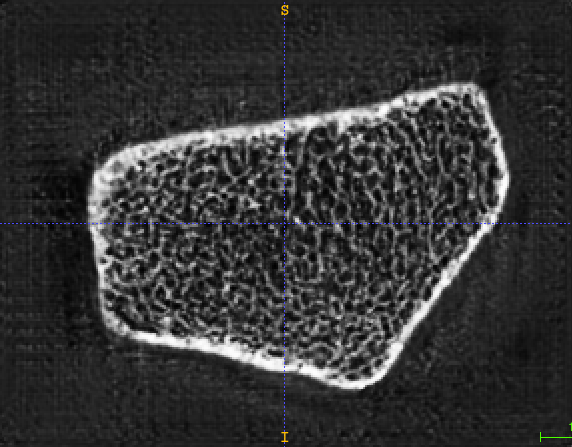}\vspace{-.3em}
		
		\hspace{.6em}
		\includegraphics[width=.162\textwidth]{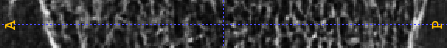}\hspace{1.38em}
		\includegraphics[width=.162\textwidth]{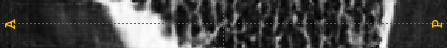}\hspace{1.38em}
		\includegraphics[width=.162\textwidth]{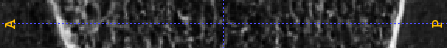}\hspace{1.38em}
		\includegraphics[width=.162\textwidth]{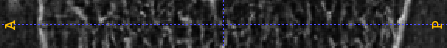}\hspace{1.38em}
		\includegraphics[width=.162\textwidth]{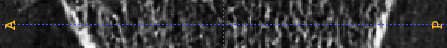}
		
		\vspace{.7cm}
		
		\includegraphics[height=.09\textheight]{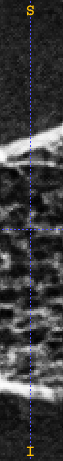}\hspace{-.1em}
		\includegraphics[height=.09\textheight]{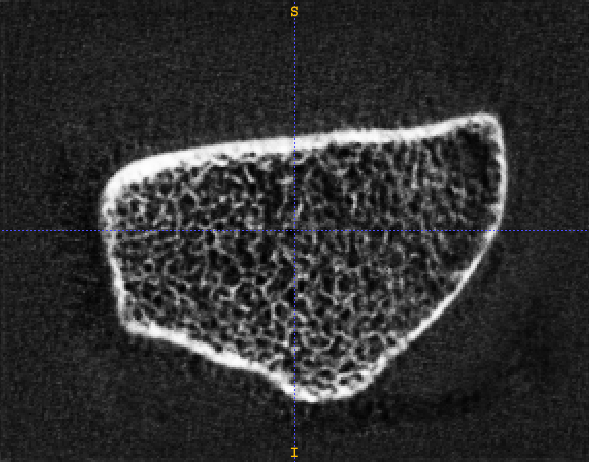}\hspace{.5em}
		\includegraphics[height=.09\textheight]{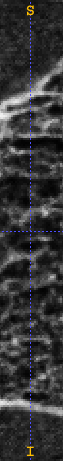}\hspace{-.1em}
		\includegraphics[height=.09\textheight]{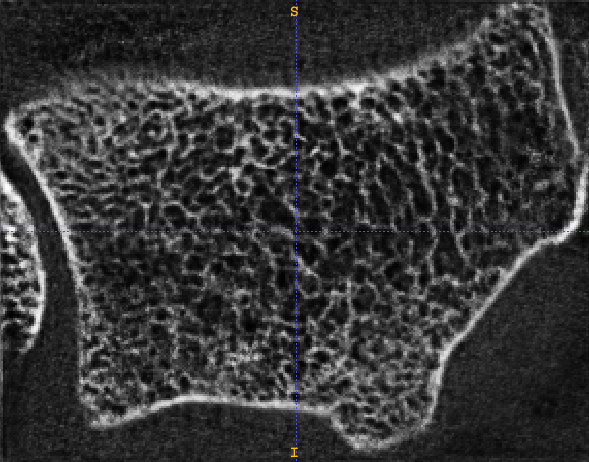}\hspace{.5em}
		\includegraphics[height=.09\textheight]{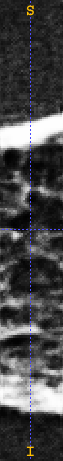}\hspace{-.1em}
		\includegraphics[height=.09\textheight]{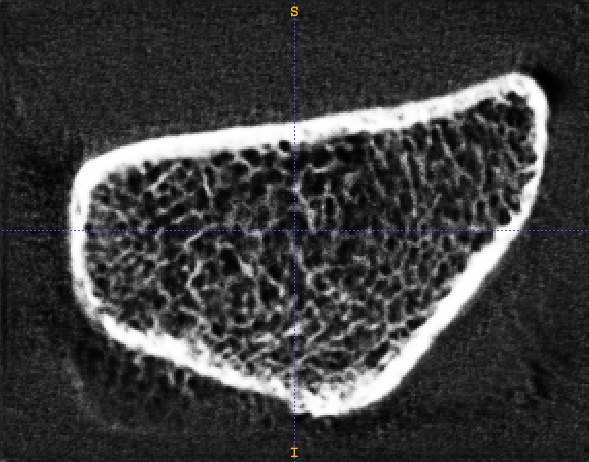}\hspace{.5em}
		\includegraphics[height=.09\textheight]{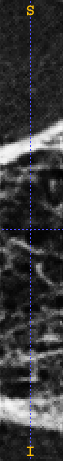}\hspace{-.1em}
		\includegraphics[height=.09\textheight]{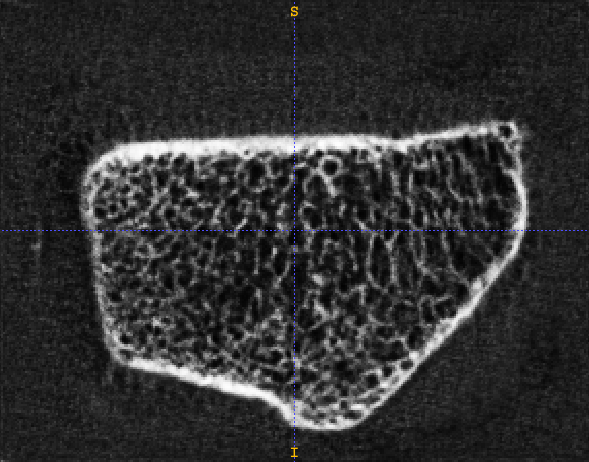}\hspace{.5em}
		\includegraphics[height=.09\textheight]{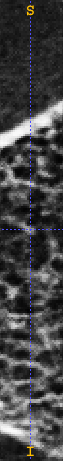}\hspace{-.1em}
		\includegraphics[height=.09\textheight]{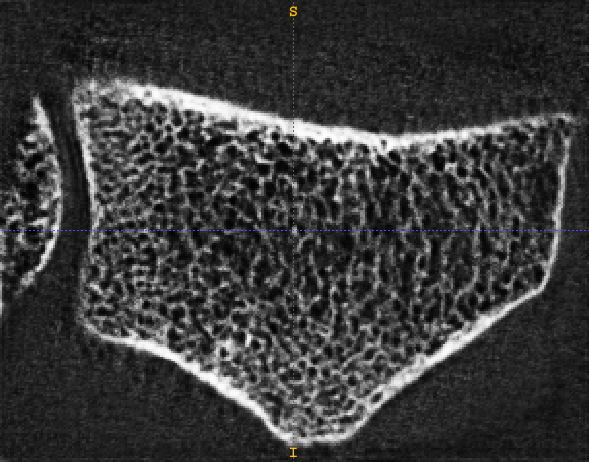}\vspace{-.3em}
		
		\hspace{.6em}
		\includegraphics[width=.162\textwidth]{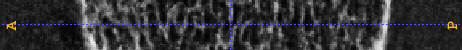}\hspace{1.38em}
		\includegraphics[width=.162\textwidth]{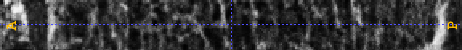}\hspace{1.38em}
		\includegraphics[width=.162\textwidth]{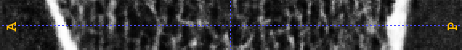}\hspace{1.38em}
		\includegraphics[width=.162\textwidth]{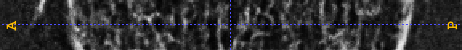}\hspace{1.38em}
		\includegraphics[width=.162\textwidth]{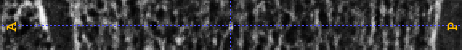}
		
		\caption{Ten HR-pQCT volumes sampled from the proposed 3D-ProGAN (first row) and 3D-StyleGAN (second row). Synthesized volumes have spatial size of \numproduct{32x288x224}.}
		\label{fig:samples}
	\end{figure}

	During training, data quality of synthesized instances is assessed after every 1000 generator updates via FID, precision and recall (cf. \ref{ssec:vali}). Results are represented from stage five with final data resolution \numproduct{32x288x224}. The truncation trick \cite{karras2017,karras2019} is deployed in Figure \ref{fig:samples}. For 3D-ProGAN a truncated normal distribution with truncation level $1.8$ is considered for sampling the latent codes. For 3D-StyleGAN, a latent code $w\in \mc W$ is normalised such that $w_\textit{norm}\coloneqq \overline{w} + \psi\cdot(w-\overline{w})$ where $\overline{w}\coloneqq \mathbb{E}_{z\sim\mc{N}(\vec{0},\text{Id})}\Phi(z)$ denotes the average latent code and $\psi$ is set to $0.8$. 
	
	\begin{wraptable}{l}{.67\columnwidth}
		\tiny
		\centering
		\caption{Quantitative results for different hyper-parameter settings. Considered hyper-parameters are truncation level ({tr}), channel size of the critic ($c_c$), channel size of the generator ($c_g$), learning rate ($\alpha$), number of critic iterations per generator updates (${n_c}$).}
		\begin{tabularx}{.65\columnwidth}{rcccc ccccc}
			
			\toprule[1.2pt]
			\textbf{tr} &$\mathbf{c_c}$& $\mathbf{c_g}$ & $\mathbf{\alpha}$ & $\mathbf{n_c}$ & \textbf{\fidinc} & \textbf{\fidres} & \textbf{\fidvgsa} &\textbf{prec} &\textbf{ rec} \\
			\midrule
			\multicolumn{5}{c}{\textbf{3D-ProGAN}}&\multicolumn{5}{c}{} \\
			\cmidrule(lr){1-5}
			\cmidrule(lr){6-10}
			5& 16 & 20 & \num{4e-3} & 5& 23.54&0.044&0.182&0.91&\textbf{0.91} \\ 
			1.8& 16 & 20 & \num{4e-3} & 5& 23.39&0.045&0.233&0.95&0.86 \\ 
			5& 12 & 20 & \num{4e-3} & 5& 25.98&0.080&0.333&0.94&0.90 \\ 
			1.8& 12 & 20 & \num{4e-3} & 5& 27.05&0.044&0.454&0.96&0.83 \\ 
			5& 20 & 20 & \num{3e-3} & 7& \textbf{21.59}&\textbf{0.040}&0.219&0.95&0.86 \\ 
			1.8& 20 & 20 & \num{3e-3} & 7& 23.31&0.259&0.274&0.97&0.82 \\
			\midrule
			\multicolumn{5}{c}{\textbf{3D-StyleGAN}}&\multicolumn{5}{c}{} \\
			\cmidrule(lr){1-5}
			\cmidrule(lr){6-10}
			1& 16 & 20 & \num{4e-3} & 6& 26.29&1.478&0.157&0.94&0.89 \\ 
			0.8& 16 & 20 & \num{4e-3} & 6& 28.99&1.343&0.258&\textbf{0.98}&0.78 \\ 
			1& 16 & 16 & \num{2e-3} & 6& 25.91&0.198&0.329&0.93&0.86 \\ 
			0.8& 16 & 16 & \num{2e-3} & 6& 28.11&0.883&0.571&0.97&0.75 \\ 
			1& 16 & 20 & \num{4e-3} & 5& 26.32&0.290&\textbf{0.151}&0.93&0.85 \\ 
			0.8& 16 & 20 & \num{4e-3} & 5& 29.07&0.509&0.206&0.96&0.70 \\
			\bottomrule[1.2pt]
		\end{tabularx}
		\label{tab:quantitative}
	\end{wraptable}

	
	Table \ref{tab:quantitative} summarizes the results for the quantitative validation metrics described in \ref{ssec:vali}. For both methods, 3D-ProGAN and 3D-StyleGAN, the three most successful runs with slightly differing hyper-parameter settings are considered for validation.  With a \fidinc\ and \fidres\ equal to 21.59 and 0.04, respectively, superior performance with respect to those two metrics is achieved by 3D-ProGAN. In terms of \fidvgsa, 3D-StyleGAN significantly outperforms 3D-ProGAN. Interestingly, 3D-StyleGAN also yields the highest precision, while in general higher recall is achieved by 3D-ProGAN.
	Indeed, comparing second row of images (as produced by 3D-StyleGAN) with first row (3D-ProGAN) in Figure \ref{fig:samples} clearly shows superiority regarding perceptual quality for 3D-StyleGAN.
	It is recommended to view the image enlarged to better observe the high-resolution quality and synthesized high-frequency details.\\
	
	It should be noted that the validation metric \fidres\ exhibits rather high variance, especially for the 3D-StyleGAN method. Arguably, due to the noise in the training data and consequently in the generated data, the features extracted by a 3D ResNet pre-trained on medical data \cite{chen2019} may not be representative.
	Further samples with varying truncation levels are displayed in Figures \ref{fig:pggan_trunc} and \ref{fig:stylegan_trunc} (see Appendix).
	
	During the evaluation process, a graphical user interface was implemented. The use of the GUI for truncation-triggered data synthesis and download is visualised in short demo videos:
	
	\begin{itemize}
		\item \url{https://www.youtube.com/watch?v=K8UbsFTSaqE} (3D-ProGAN)
		\item \url{https://www.youtube.com/watch?v=4VPDUZ3Pbk8} (3D-StyleGAN)
	\end{itemize}

	\subsection{Image Transition}
	The previous section demonstrates the ability to successfully generate high-resolution bone CTs with high diversity. Generation by sampling latent codes can be used to extend data sets in an unconditioned manner. In this case, the distribution of a given attribute in the synthesised data is very likely to follow the distribution of the same attribute in the training set. In this section, a method is proposed for synthesising data with respect to a particular attribute.
	
	Image transition aims to semantically interpolate two medical samples by propagating a weighted sum of the corresponding latent codes through a fixed generative function. This is suitable for investigating the plausibility of the inverted codes -- for a good GAN inversion, the spatial and semantic attributes should vary continuously during the transition from one inverted code to the other inverted counterpart. If the underlying scans of both codes share a certain attribute, all generated scans during the transition should also share this attribute.\\
	
	For investigation, two specific properties of bone HR-pQCT data are targeted --  trabecular bone mineral density (Tb.BMD) and cortical bone mineral density (Ct.BMD).
	Ct.BMD and Tb.BMD correspond to the average mineral density (i.e. X-ray beam attenuation) within the voxel volume of the cortical and trabecular compartments, respectively, and is calculated directly from the gray-scale image data \cite{whittier2020}. These attributes have been shown to be statistically linked to bone fracture risk \cite{whittier2023}. 
	As the training data is comprised of images from patients who experienced a bone fracture, the distribution of Ct.BMD and Tb.BMD values in the data set is not normally distributed, exhibiting a slight bias.\\
	
	\begin{figure}[thb!]
		\includegraphics[height=.09\textheight]{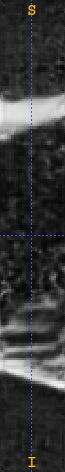}\hspace{-.1em}
		\includegraphics[height=.09\textheight]{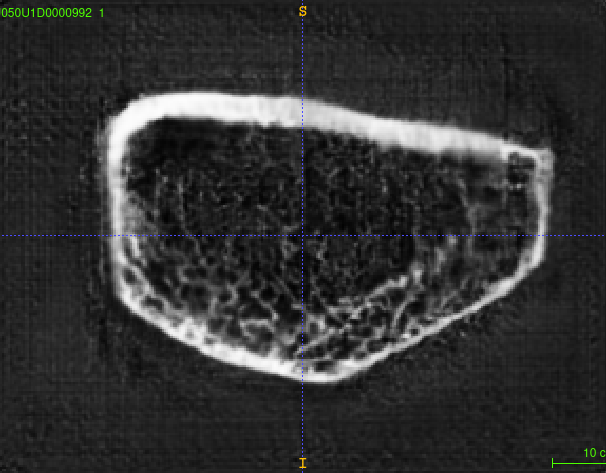}\hspace{.5em}
		\includegraphics[height=.09\textheight]{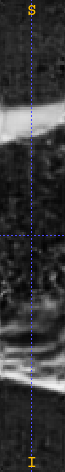}\hspace{-.1em}
		\includegraphics[height=.09\textheight]{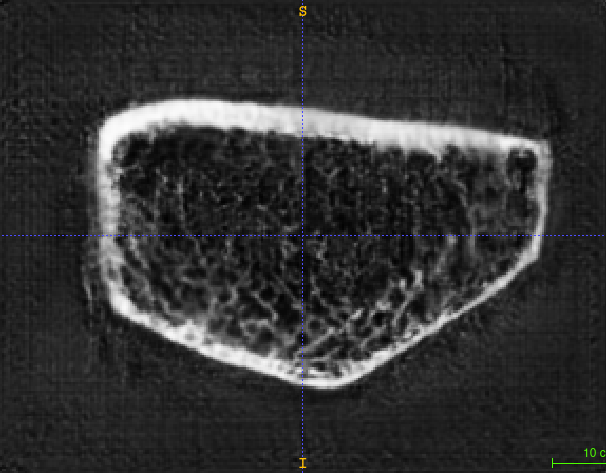}\hspace{.5em}
		\includegraphics[height=.09\textheight]{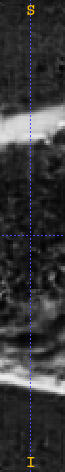}\hspace{-.1em}
		\includegraphics[height=.09\textheight]{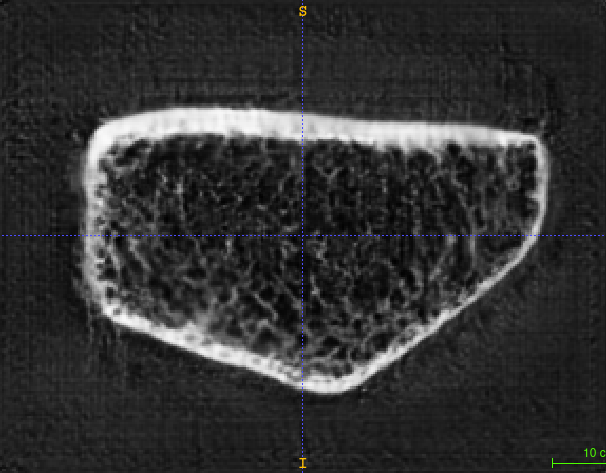}\hspace{.5em}
		\includegraphics[height=.09\textheight]{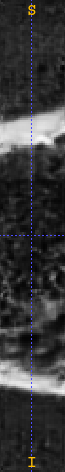}\hspace{-.1em}
		\includegraphics[height=.09\textheight]{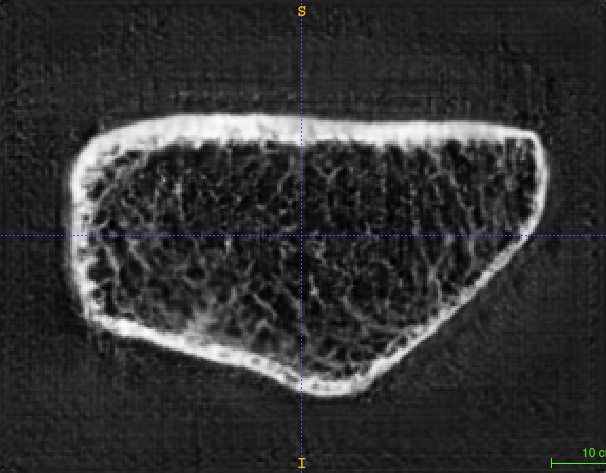}\hspace{.5em}
		\includegraphics[height=.09\textheight]{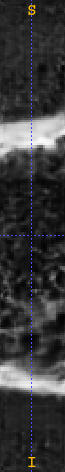}\hspace{-.1em}
		\includegraphics[height=.09\textheight]{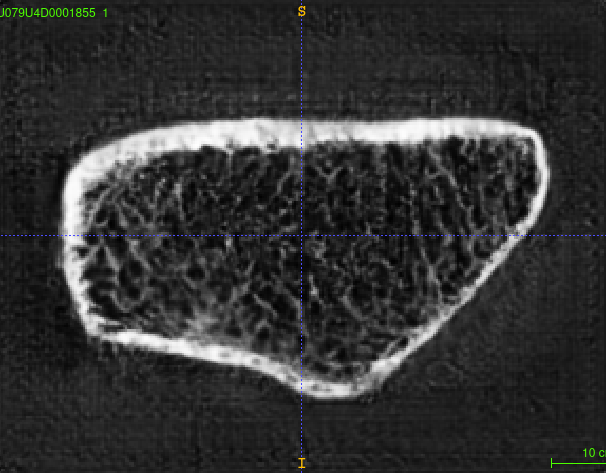}\vspace{-.4em}
		
		\hspace{.6em}
		\includegraphics[width=.162\textwidth]{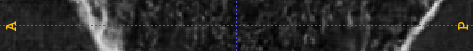}\hspace{1.38em}
		\includegraphics[width=.162\textwidth]{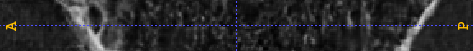}\hspace{1.38em}
		\includegraphics[width=.162\textwidth]{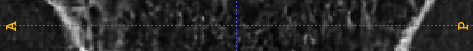}\hspace{1.38em}
		\includegraphics[width=.162\textwidth]{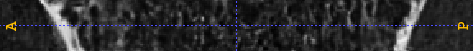}\hspace{1.38em}
		\includegraphics[width=.162\textwidth]{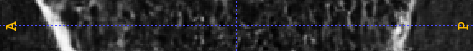}
		
		\vspace{.7cm}
		\includegraphics[height=.09\textheight]{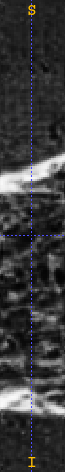}\hspace{-.1em}
		\includegraphics[height=.09\textheight]{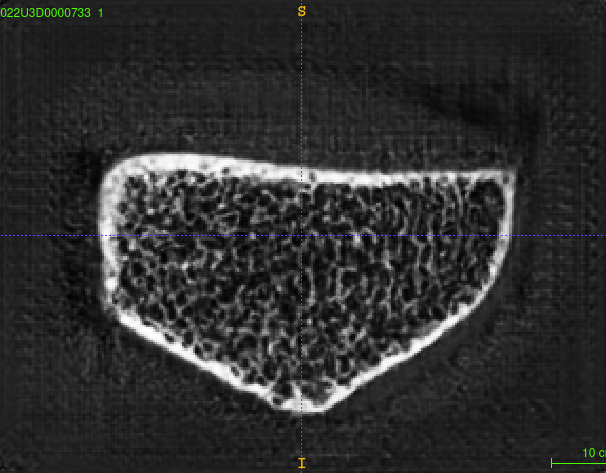}\hspace{.5em}
		\includegraphics[height=.09\textheight]{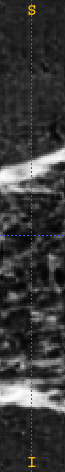}\hspace{-.1em}
		\includegraphics[height=.09\textheight]{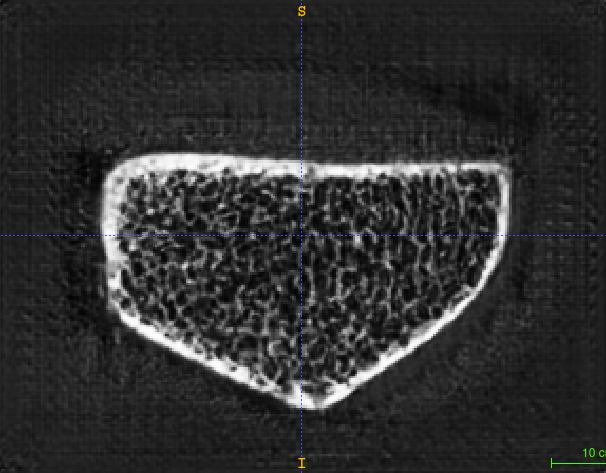}\hspace{.5em}
		\includegraphics[height=.09\textheight]{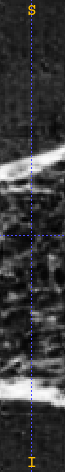}\hspace{-.1em}
		\includegraphics[height=.09\textheight]{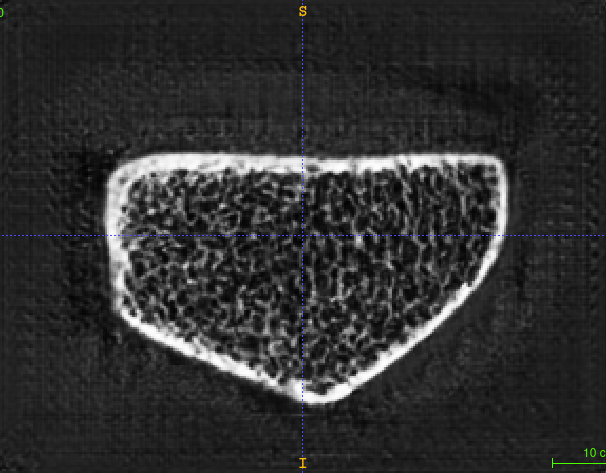}\hspace{.5em}
		\includegraphics[height=.09\textheight]{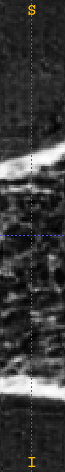}\hspace{-.1em}
		\includegraphics[height=.09\textheight]{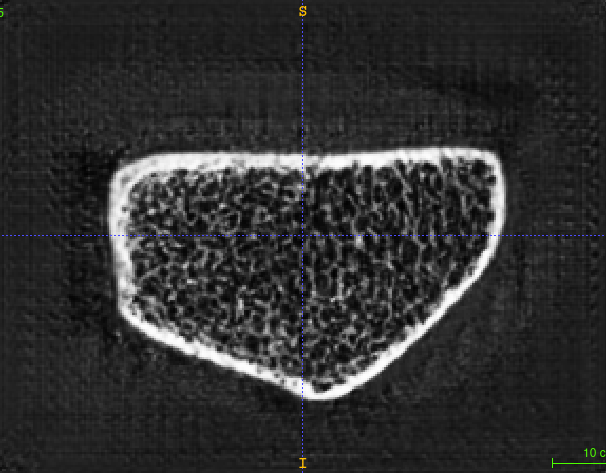}\hspace{.5em}
		\includegraphics[height=.09\textheight]{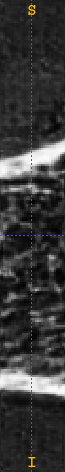}\hspace{-.1em}
		\includegraphics[height=.09\textheight]{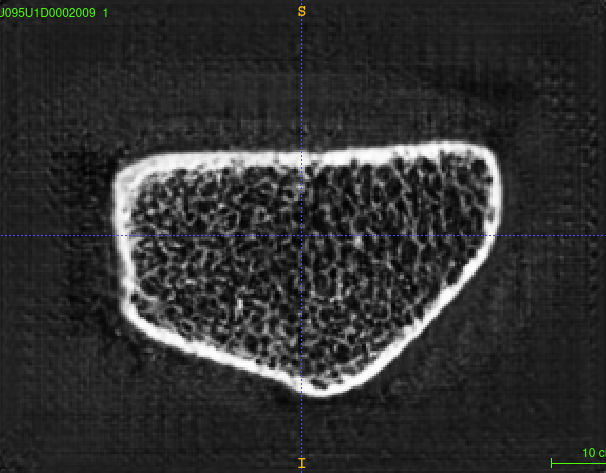}\vspace{-.4em}
		
		\hspace{.6em}
		\includegraphics[width=.162\textwidth]{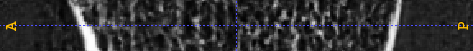}\hspace{1.38em}
		\includegraphics[width=.162\textwidth]{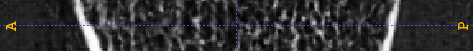}\hspace{1.38em}
		\includegraphics[width=.162\textwidth]{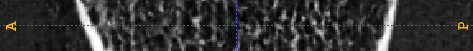}\hspace{1.38em}
		\includegraphics[width=.162\textwidth]{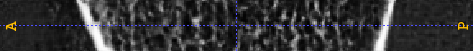}\hspace{1.38em}
		\includegraphics[width=.162\textwidth]{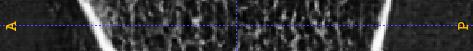}
		
		\caption{First row: samples with weak trabecular bone mineralization (Tb.BMD). Second row:        samples with weak cortical bone mineralization (Ct.BMD). From left to right:                        $x_1,\ x_{1,2}^{0.25},\ x_{1,2}^{0.5},\ x_{1,2}^{0.75},\ x_2$.}
		\label{fig:transition}
	\end{figure}
	
	Let $x_1,x_2 \in \mc X$ denote two samples from the training set with a small value for Tb.BMD. The GAN inversion strategy discussed in \ref{ssec:inversion} is applied for 3D-ProGAN. According to \eqref{eq:min}, this yields $z_1 \coloneqq z_\text{opt}(x_1)$ and $z_2 \coloneqq z_\text{opt}(x_2)$. During transition, the generative function $G$ of 3D-ProGAN is used to generate new samples $x_{1,2}^\alpha = G\left(\alpha\cdot z_1+(1-\alpha)\cdot z_2\right)$ for $\alpha \in [0,1]$. In Figure \ref{fig:transition}, the generated results
	are displayed in the first row. Obviously, the average bone mineralization in the trabecular compartment is weak for all scans, while a smooth spatial transition from $x_1$ to $x_2$ can be observed. The second row shows the same procedure repeated with samples for $x_1$ and $x_2$ exhibiting small Ct.BMD values. The implemented GUI provides an interactive way to use the image transition to synthesise new data for augmentation. A demonstration video can be found here:\\
	\url{https://www.youtube.com/watch?v=j6Fh0a4r1Rw}.

	\subsection{Style Mixing}
	The interpolation between scans discussed above allows for a smooth transition between different shapes while preserving certain attributes. However, it is also possible to fix a certain property of the first patient (e.g. shape) and mix it with the given style of a second patient (e.g. trabecular properties). The 3D-StyleGAN allows manipulation of the output of the generative function by using the style transfer capability of the network, where two latent codes of the learned latent space $\mc W$ are included in the generation process. As described in \ref{ssec:arch}, a latent code $w$ is converted  by learned affine transformations into 15 different style codes, which are fed into the generative function using weight demodulation. The idea of style mixing is to feed the style codes based on the source scan and the codes from the target scan to the generator.
	
	\begin{figure}[thb!]
		
		\hspace{3.68cm}
		\includegraphics[height=.11\textheight]{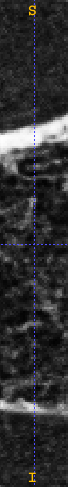}\hspace{-.05cm}
		\includegraphics[height=.11\textheight]{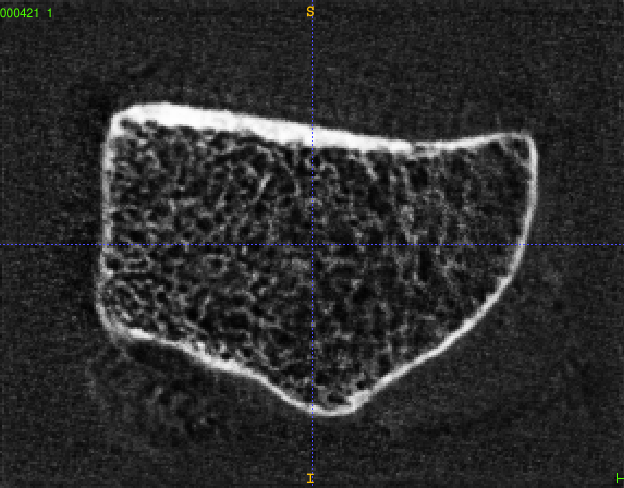}\hspace{.15cm}
		\includegraphics[height=.11\textheight]{figs_arxiv/style_mixing/s0_02.png}\hspace{-.05cm}
		\includegraphics[height=.11\textheight]{figs_arxiv/style_mixing/s0_01.png}\hspace{.15cm}
		\includegraphics[height=.11\textheight]{figs_arxiv/style_mixing/s0_02.png}\hspace{-.05cm}
		\includegraphics[height=.11\textheight]{figs_arxiv/style_mixing/s0_01.png}\vspace{-0.07cm}
		
		\hspace{4.09cm}
		\includegraphics[width=.197\textwidth]{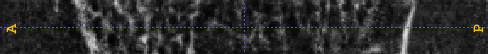}\hspace{.583cm}
		\includegraphics[width=.197\textwidth]{figs_arxiv/style_mixing/s0_00.png}\hspace{.583cm}
		\includegraphics[width=.197\textwidth]{figs_arxiv/style_mixing/s0_00.png}\vspace{.2cm}
		
		\includegraphics[height=.11\textheight]{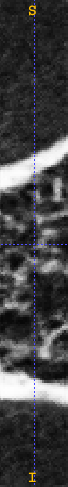}\hspace{-.05cm}
		\includegraphics[height=.11\textheight]{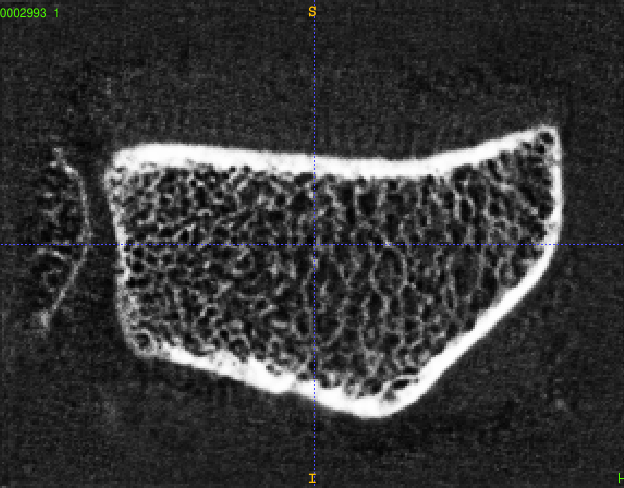}\hspace{.15cm}
		\includegraphics[height=.11\textheight]{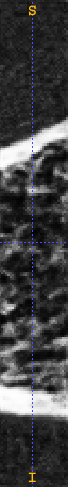}\hspace{-.05cm}
		\includegraphics[height=.11\textheight]{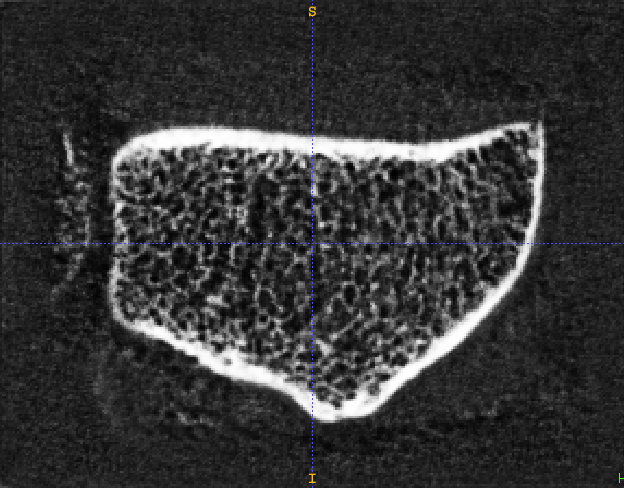}\hspace{.15cm}
		\includegraphics[height=.11\textheight]{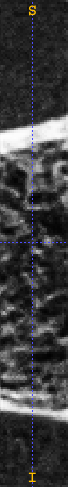}\hspace{-.05cm}
		\includegraphics[height=.11\textheight]{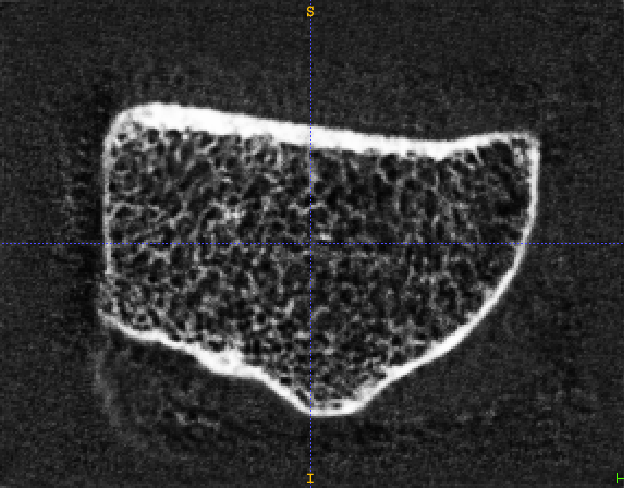}\hspace{.15cm}
		\includegraphics[height=.11\textheight]{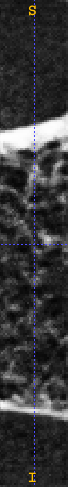}\hspace{-.05cm}
		\includegraphics[height=.11\textheight]{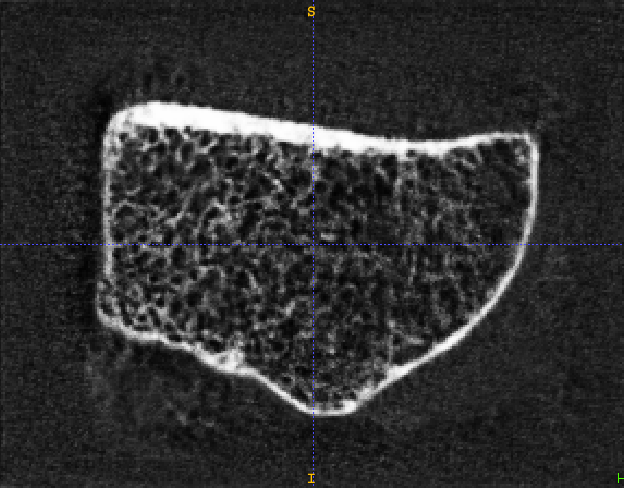}\vspace{-0.07cm}
		
		\hspace{0.295cm}
		\includegraphics[width=.197\textwidth]{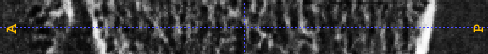}\hspace{.583cm}
		\includegraphics[width=.197\textwidth]{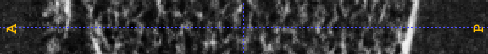}\hspace{.583cm}
		\includegraphics[width=.197\textwidth]{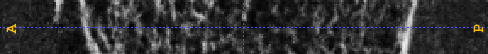}\hspace{.583cm}
		\includegraphics[width=.197\textwidth]{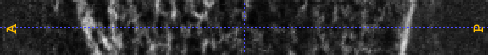}

		\vspace{1.5cm}
		\hspace{3.68cm}
		\includegraphics[height=.11\textheight]{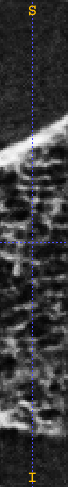}\hspace{-.05cm}
		\includegraphics[height=.11\textheight]{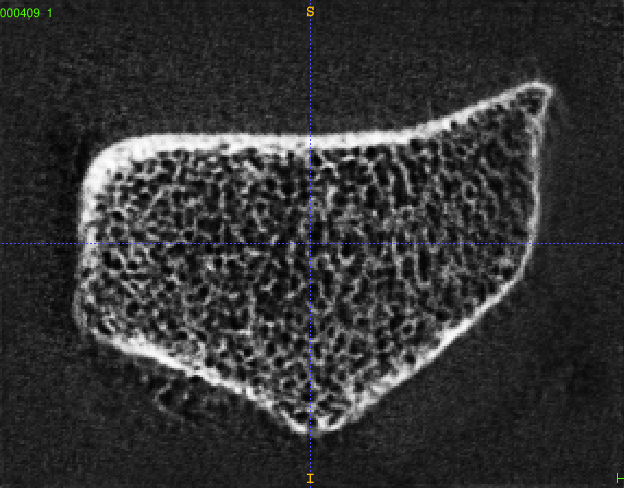}\hspace{.15cm}
		\includegraphics[height=.11\textheight]{figs_arxiv/style_mixing/s1_02.png}\hspace{-.05cm}
		\includegraphics[height=.11\textheight]{figs_arxiv/style_mixing/s1_01.png}\hspace{.15cm}
		\includegraphics[height=.11\textheight]{figs_arxiv/style_mixing/s1_02.png}\hspace{-.05cm}
		\includegraphics[height=.11\textheight]{figs_arxiv/style_mixing/s1_01.png}\vspace{-0.07cm}
		
		\hspace{4.09cm}
		\includegraphics[width=.197\textwidth]{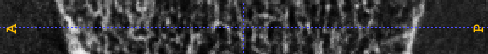}\hspace{.583cm}
		\includegraphics[width=.197\textwidth]{figs_arxiv/style_mixing/s1_00.png}\hspace{.583cm}
		\includegraphics[width=.197\textwidth]{figs_arxiv/style_mixing/s1_00.png}\vspace{.2cm}
		
		\includegraphics[height=.11\textheight]{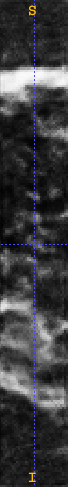}\hspace{-.05cm}
		\includegraphics[height=.11\textheight]{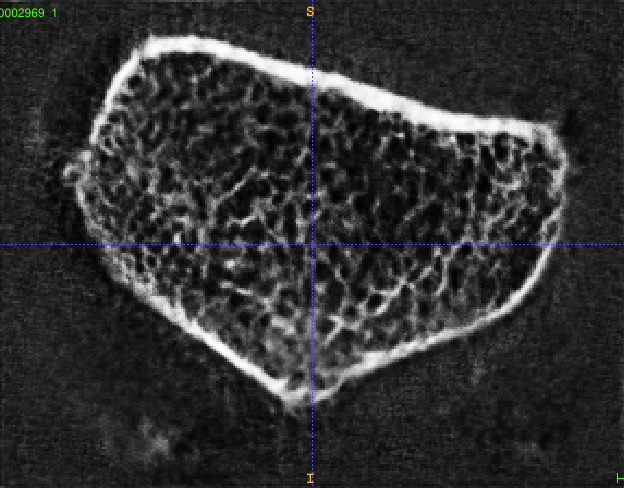}\hspace{.15cm}
		\includegraphics[height=.11\textheight]{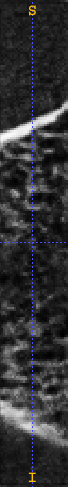}\hspace{-.05cm}
		\includegraphics[height=.11\textheight]{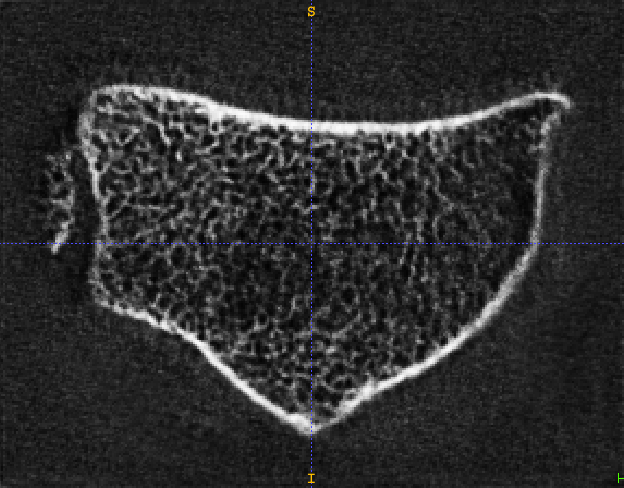}\hspace{.15cm}
		\includegraphics[height=.11\textheight]{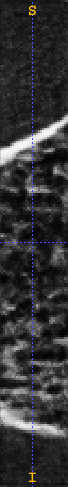}\hspace{-.05cm}
		\includegraphics[height=.11\textheight]{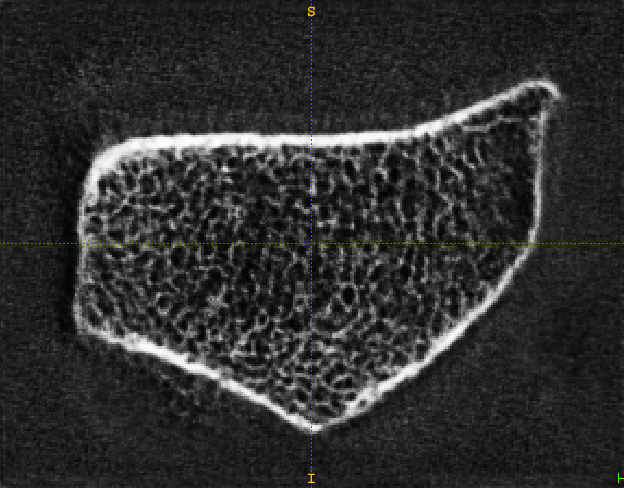}\hspace{.15cm}
		\includegraphics[height=.11\textheight]{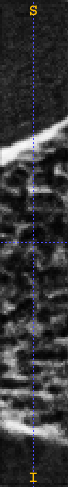}\hspace{-.05cm}
		\includegraphics[height=.11\textheight]{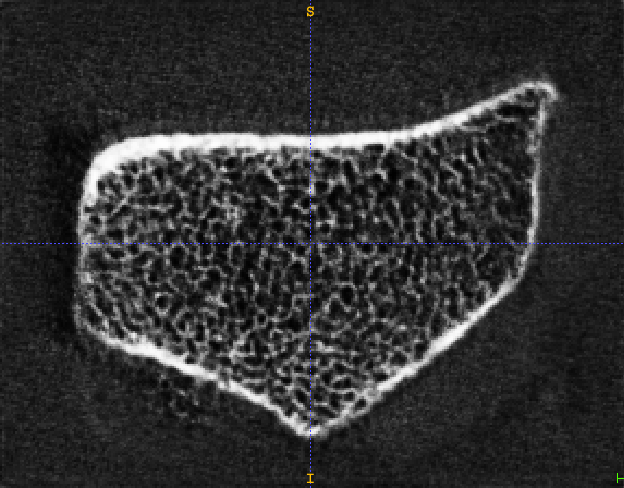}\vspace{-0.07cm}
		
		\hspace{0.295cm}
		\includegraphics[width=.197\textwidth]{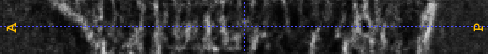}\hspace{.583cm}
		\includegraphics[width=.197\textwidth]{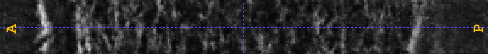}\hspace{.583cm}
		\includegraphics[width=.197\textwidth]{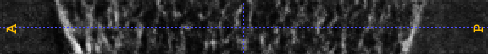}\hspace{.583cm}
		\includegraphics[width=.197\textwidth]{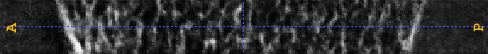}
		
		\caption{An illustration of the style combination based on the 3D-StyleGAN approach. For both examples, the first row denotes the source image (real patient data). The second row contains the target image at the left most position and style mix results where the style of the source is fed to the generator in the first three convolutional layers, in the first seven layers and in the first twelve layers (from left to right).}
		\label{fig:style_mix}
	\end{figure}
	
	Let $s\in \mc X$ and $t \in \mc X$ denote the source and target images of real patients, respectively. Applying the GAN inversion strategy in \ref{ssec:inversion} for 3D-StyleGAN yields $w_s \coloneqq w_{\text{opt}}(s)$ and $w_t \coloneqq w_\text{opt}(t)$, where both inverted codes are forced to reside in $\mc W$ by the latent discriminator (cf. \eqref{eq:lat_disc}). 3D-StyleGAN consists of a generator $\tilde G :\mc{W}^{15}\to\mc X$ that takes 15 different style vectors based on latent input $w$ and feeds them to the convolutional layers by weight demodulation \cite{karras2019}. Variation of different styles is enabled by using style vectors based on both, latent source code $w_s$ and latent target code $w_t$. Let $x_{s,t}^a$ denote a generated sample of 3D-StyleGAN that used style vectors of $w_s$ for the first $a$ convolution layers and style vectors of $w_t$ for the remaining convolution layers.\\
	
	Figure \ref{fig:style_mix} shows sample results for this technique. The top-most row shows the same source image three times, taken from a patient with a comparatively low Ct.BMD value. The second row displays the target image with a high Ct.BMD value as well as the style mix results $x_{s,t}^3$, $x_{s,t}^7$ and $x_{s,t}^{12}$. It can be observed that $x_{s,t}^3$ yields an interpolation of both shapes and a strong cortical bone structure. Increasing the numbers of source style vectors to seven in $x_{s,t}^7$ yields a bone with similar shape to the source but with the cortical property of the target. This is an essential result for this study -- it is possible to apply a certain attribute from a target image to the shape of another source. Only using three style vectors of the target scan in the last three convolution layers ($x_{s,t}^{12}$) yields nearly no differences to the source scan.
	
	The third and fourth row of figure \ref{fig:style_mix} show the mix approach repeated for trabecular bone mineral density. Again, $x_{s,t}^3$ shows a transition between both shapes with small Tb.BMD value, $x_{s,t}^7$ yields a copy of the source image with significant changes in the trabecular structure, and $x_{s,t}^{12}$ is quite similar to the source image. \\
	
	In conclusion, the use of the proposed 3D-StyleGAN for style mixing appears to be another reliable tool for editing HR-pQCT attributes. It can be concluded that styles applied to low resolution convolution layers determine spatial attributes of the bone, while codes applied to higher resolution layers are responsible for variation in semantic features such as cortical or trabecular condition.

	\subsection{Attribute Editing}
	The previous section demonstrated the impact of the latent representation on different resolutions in the generative function. To complete the analysis of the relationship between latent and image space, the following section examines the interpretability of latent space. In two-dimensional applications, generative networks have been shown to automatically learn to represent multiple interpretable attributes in latent space \cite{xia2022,shen2021,shen2020}. 
	These works suggest to identify a semantically meaningful direction $n \in \R^{512}$ in order to achieve a manipulation $x_\text{edit} = G\left(z_\text{opt}(x)+\alpha n\right)$.
	
	According to an extensive literature research, this is the first study to leverage exploration of meaningful directions to the 3D case in a unsupervised manner. The approach in \cite{shen2021} is used to find the optimal direction $n^*$:
	
	\begin{equation}
		\label{eq:eigenvec}
		n^* = \argmax_{\{n\in \R^{512}\mid n^Tn=1\}} \norm{An}^2_2.
	\end{equation}
	
	The matrix $A$ denotes either the first linear layer in 3D-ProGAN, or the concatenation of 15 linear layers in 3D-StyleGAN, which convert the latent code into a style code. The term \textit{optimal} directions corresponds to a vector that causes large variations after projection by $A$. Similar to \cite{shen2021}, the top four directions $n_1,n_2,n_3,n_4$ are determined by using the eigenvectors of $A^TA$ associated with the four largest eigenvalues.\\
	
	\begin{figure}[thb!]
		\includegraphics[width=.47\textwidth]{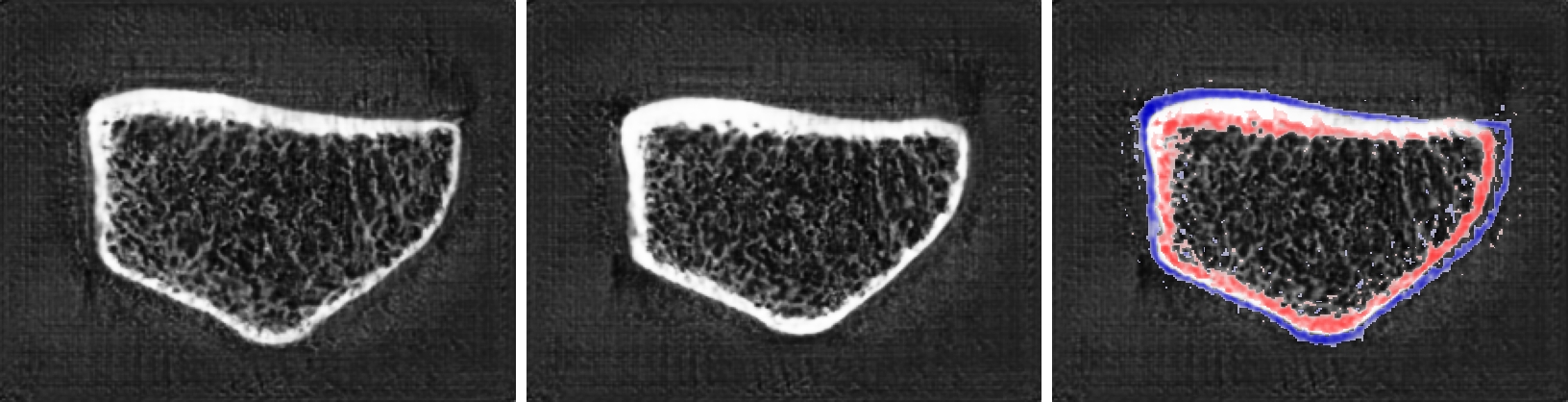}\hspace{1cm}
		\includegraphics[width=.47\textwidth]{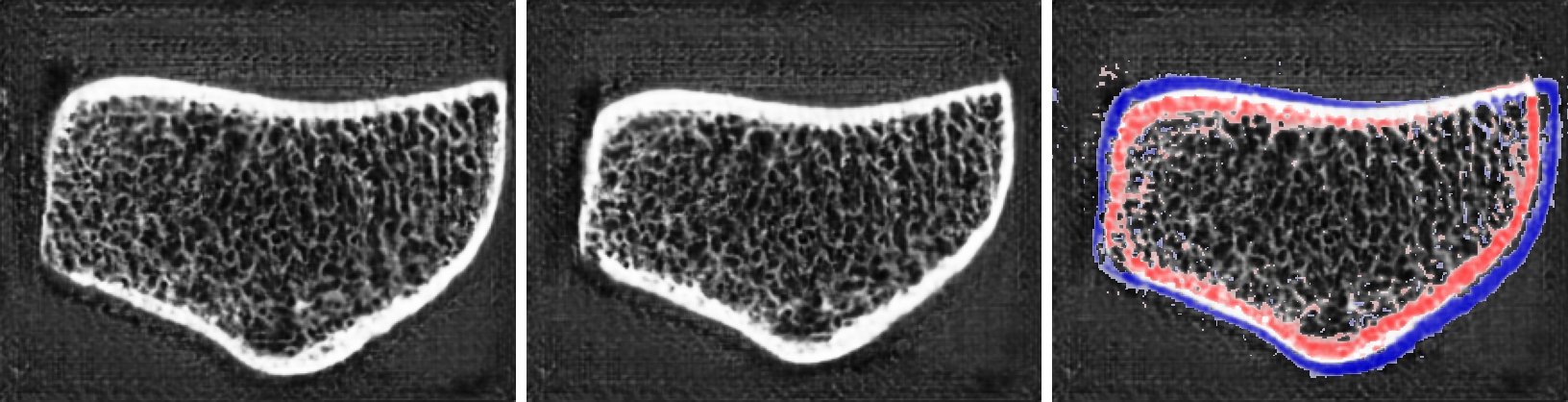}
		
		\includegraphics[width=.47\textwidth]{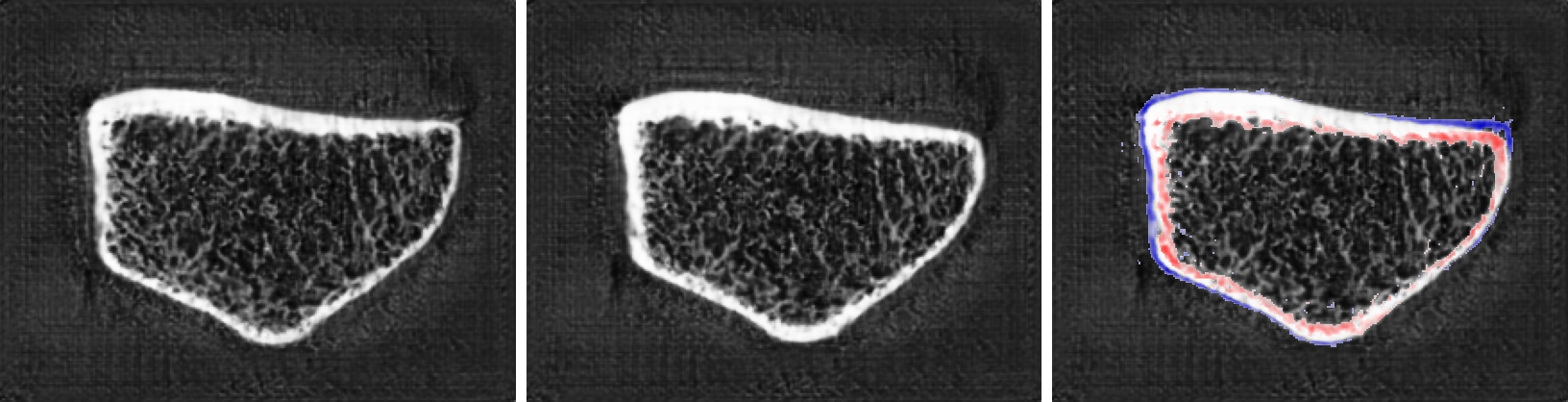}\hspace{1cm}
		\includegraphics[width=.47\textwidth]{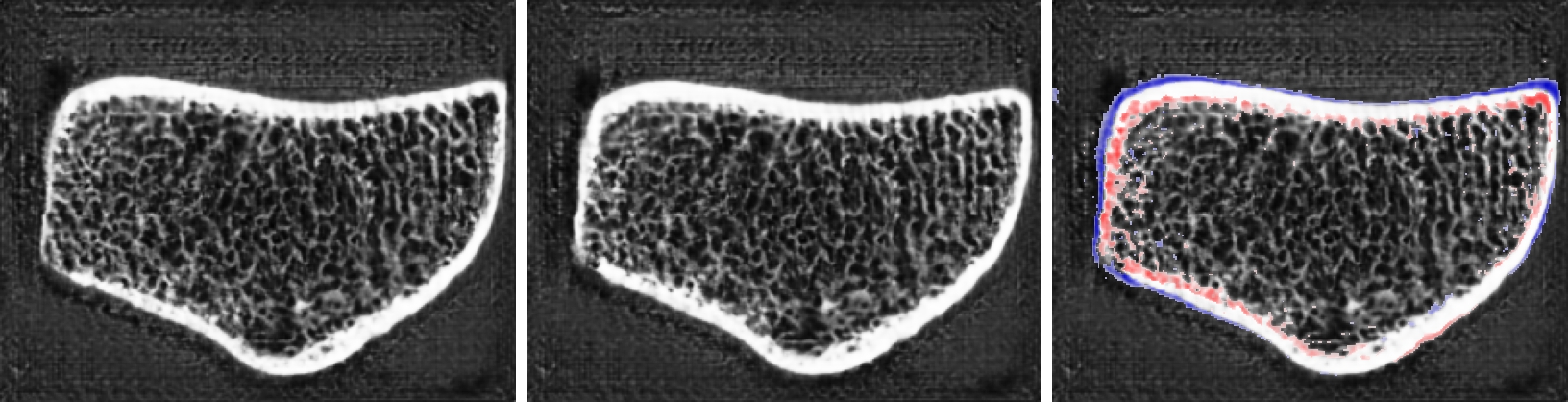}
		
		\includegraphics[width=.47\textwidth]{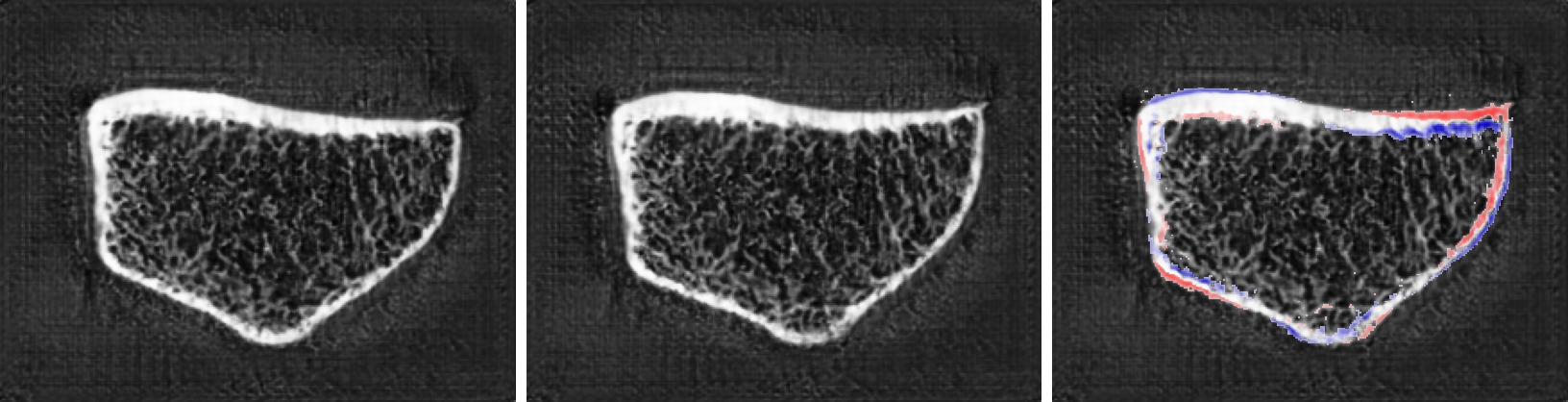}\hspace{1cm}
		\includegraphics[width=.47\textwidth]{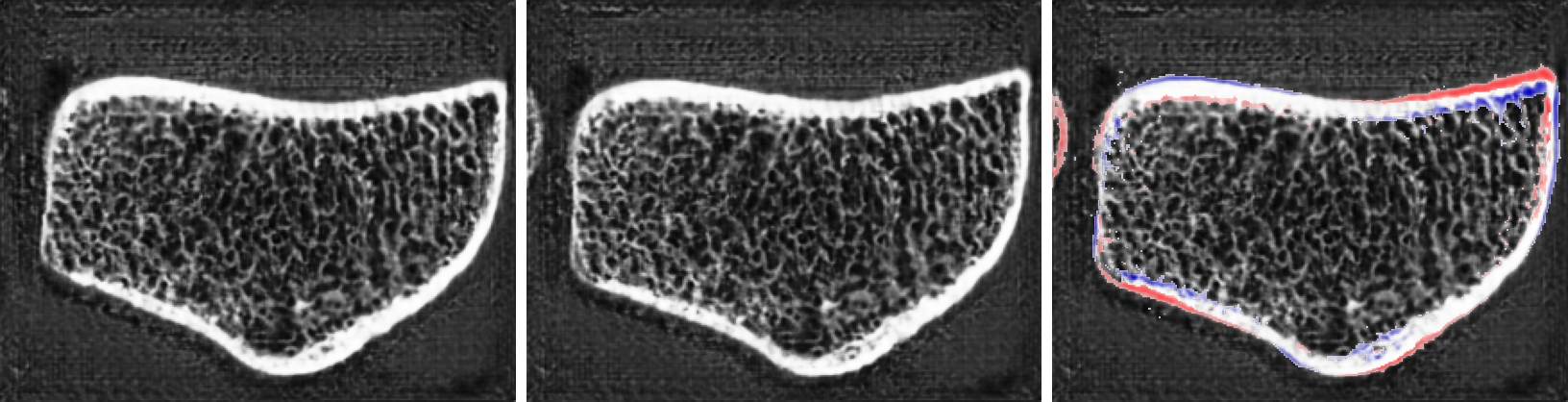}
		
		\includegraphics[width=.47\textwidth]{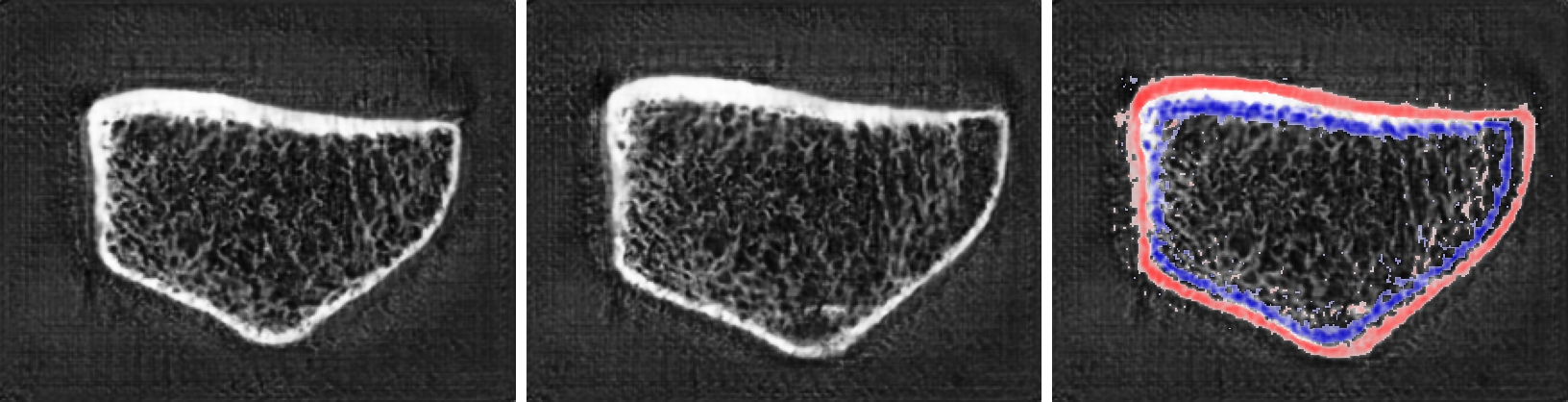}\hspace{1cm}
		\includegraphics[width=.47\textwidth]{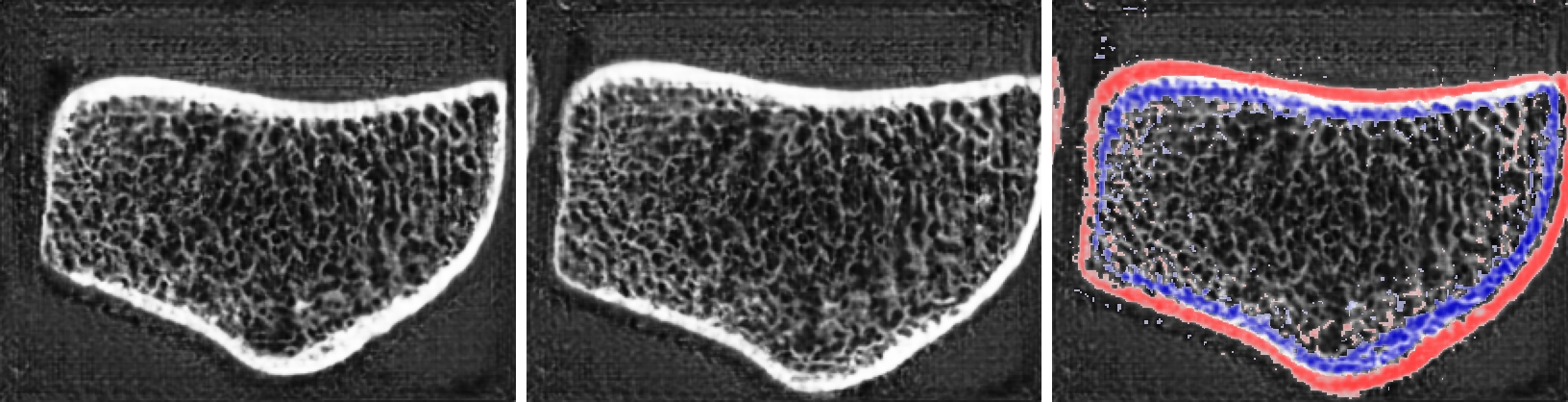}
		\caption{3D-ProGAN results for attribute editing. For each volumetric sample the center axial slice is visualized. Left: Existing patient $x$. Middle: Generated samples $G_1(z_\text{opt}(x)+\alpha n_k),\ k = 1,2,3,4$. Right: difference $G_1(z_\text{opt}(x)+\alpha n_k)-x$, where red and blue voxels denote positive and negative residuals, respectively.}
		\label{fig:pggan_attr}
	\end{figure}	
	
	Figure \ref{fig:pggan_attr} shows the latent space analysis applied to 3D-ProGAN. A subsequent analysis of the manipulated images is necessary to understand which property each direction $n_1,n_2,n_3,n_4$ encodes. The first direction $n_1$ shrinks the circumference of the cortical compartment while leaving semantic properties unchanged (first row). $n_2$ significantly enlarges the cortical compartment (second row). The third vector $n_3$ results in a slight rotation of the bone while $n_4$, complementary to $n_1$, enlarges the circumference of the cortical compartment (third and fourth row). In all editing operations, the strength of manipulation $\alpha$ equals 4. All four directions may be used in data augmentation scenarios to increase bone size, change the cortical thickness or rotate the sample. Interestingly, none of the four latent directions has a crucial impact on the trabecular properties. These may be varied by the use of eigenvector associated with smaller eigenvalues.

	\subsection{Expert Validation}
	
\begin{figure}[thb!]
		\centering
		\includegraphics[width=.32\columnwidth]{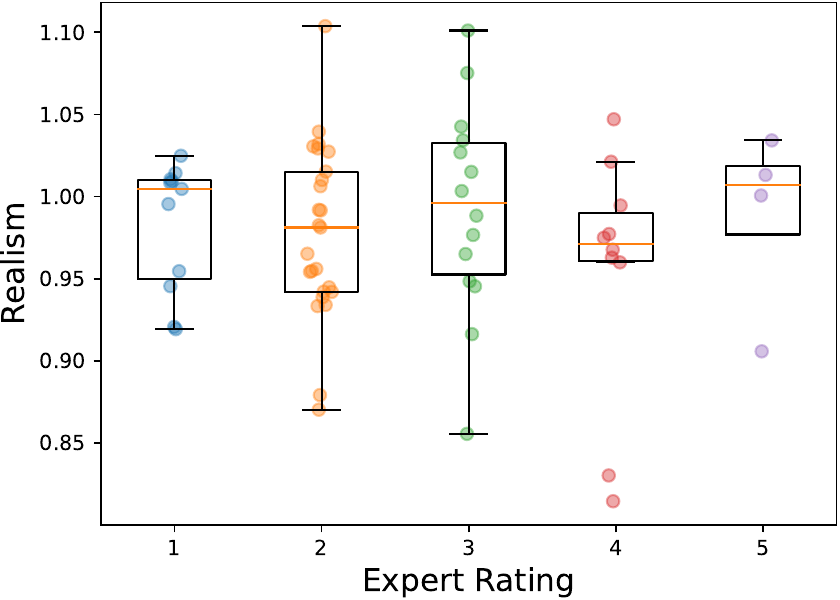}\hfill
		\includegraphics[width=.32\columnwidth]{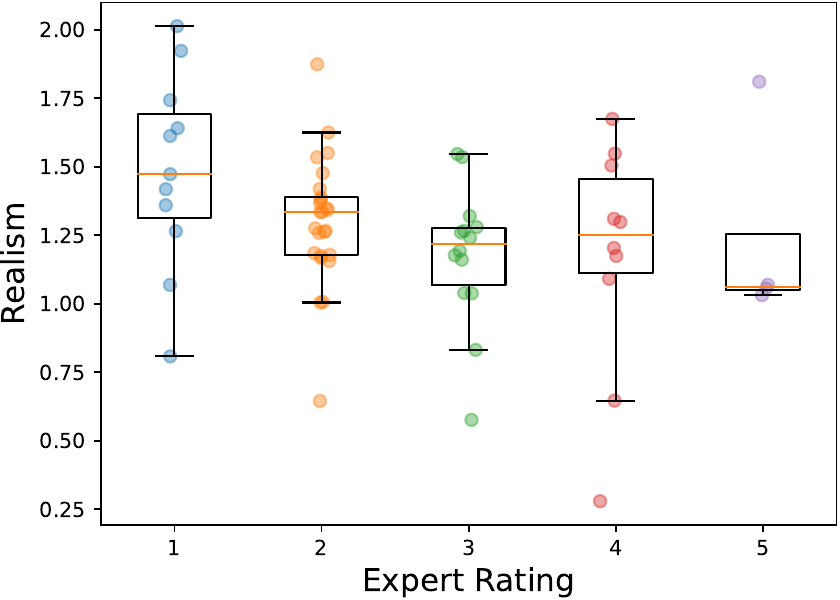}\hfill
		\includegraphics[width=.32\columnwidth]{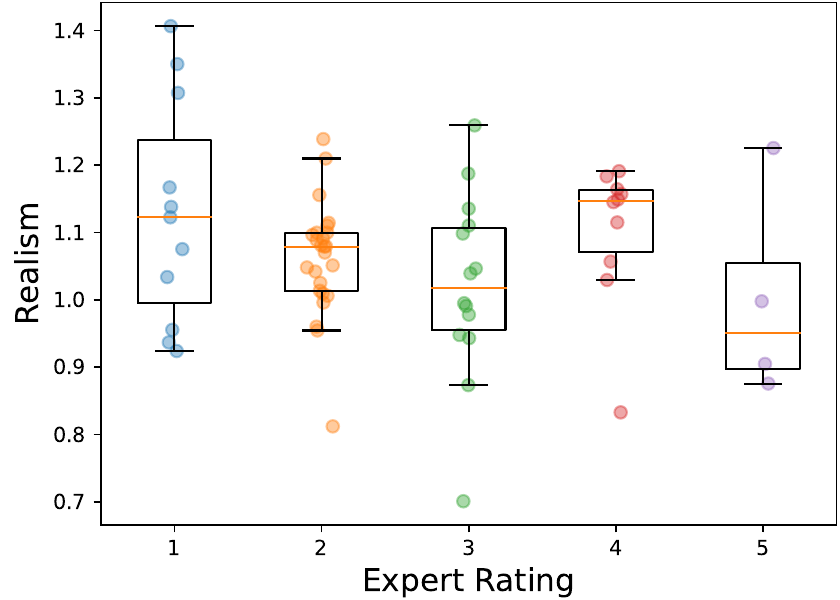}\vspace{1em}
		\includegraphics[width=.32\columnwidth]{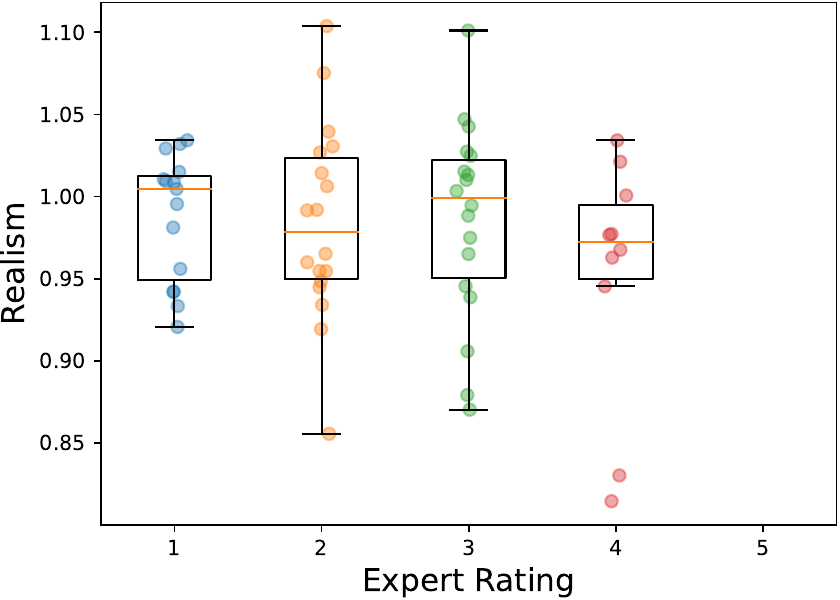}\hfill
		\includegraphics[width=.32\columnwidth]{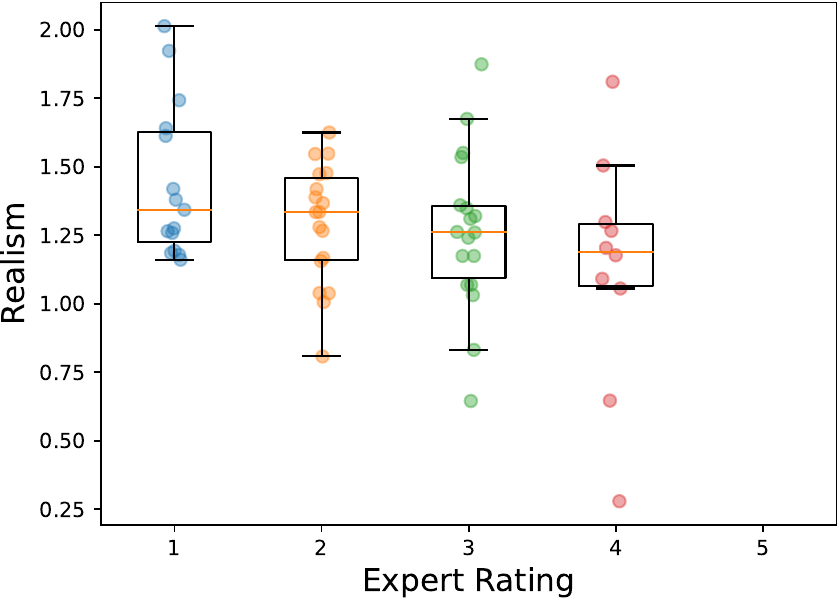}\hfill
		\includegraphics[width=.32\columnwidth]{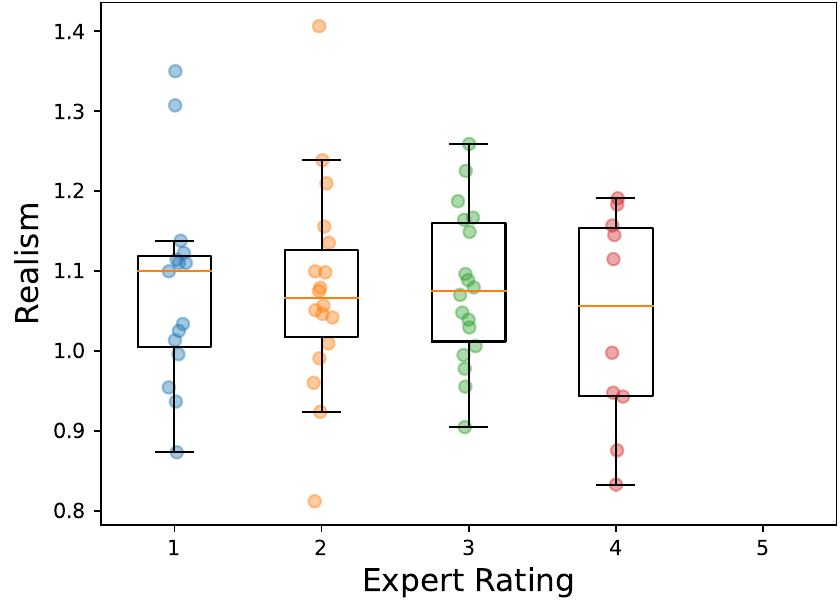}
		\caption{Comparison between computer-based realism scores and the subjective rating by expert 1 (first row) and expert 2 (second row) on HR-pQCT images. The horizontal axes denote the expert rating 1-5 while the vertical axes shows the calculated realism scores. From left to right: \rinc, \rres, \rvgsa.}
		\label{fig:realism}
	\end{figure}

	An essential research goal for this work is to investigate computer-based metrics and their ability to approximate the visual perception of human experts in the field. As already thoroughly discussed in \ref{ssec:vali}, three realism scores, based on three different feature extraction methods, are utilized: \rinc, \rres~ and \rvgsa. In contrast to the Frech\'et Inception Distance, which measures the distance between distributions quantitatively, these realism scores enable the evaluation of perceptual quality for a single sample. These metrics are evaluated using 64 synthetic volumetric images generated using the 3D-ProGAN method. These examples were also evaluated by two CT imaging experts, focusing in particular on image sharpness, valid image area, artefacts, contours and repetitive patterns in the trabecular structure. Based on these criteria, a score of 1 to 5 was assigned, with a lower score indicating a better rating. The results are depicted in Figure \ref{fig:realism}.\\
	
	No clear correlation can be found between the expert's opinion and the realism score based on the Inception v3 classification network \rinc. This is not the case for \rres~ and \rvgsa. Both realism scores are able to distinguish between low and high expert rated samples to some extent. Especially for \rres, which was generated using a 3D ResNet model pre-trained on medical data for feature extraction, the correlation is quite clear for expert 1 and 2. However, none of the considered realism metrics is able to accurately reflect the subjective opinion of a human expert. Evaluation of a larger synthetic cohort, involvement of more experts, and a wider range of feature extraction methods will be part of future research.
	
	
	\section{Conclusion \& Future Impact}
	This work demonstrates that three-dimensional generative models can be successfully trained to generate high resolution medical images of fine-detailed micro-architectures on voxel basis. In particular, progressive growing and style-based GAN architectures were shown to be viable for the synthetic creation of realistic volumetric grey-scale images. Furthermore, GAN inversion techniques are used to map measurable image attributes to directions in a low-dimensional latent space, which allows generated images to be parameterized with regard to those attributes. Considering style-based generation, it is possible to mix the characteristics of two source images, creating realistic results which combine selected properties in a controllable manner. Given the modest number of images used in training when compared with the volumes used for similar (2D) image generator networks, the results are definitely impressive. While tell-tale artefacts in the background noise are easily spotted by human experts, the overall structure and small-scale details of the generated bones closely follow the natural patterns. Regarding the naturalism, the variation of the shape outlines is, in general, very realistic and shows great variability.
	
	Naturally, this work still has some limitations. For one, the implementation of an automated realism assessment that mimics the perception of human experts mainly depends on an appropriate feature extraction method. While this study has shown that commonly used feature extraction models only approximate human perception to a certain extent, appropriate feature computation still requires further research. While an automated realism score would be greatly helpful for large batch image generation jobs, it does not impact the overall usefulness of the generative models.
	It is also to be noted that the resolution of the generated images, while already high for the standards of generative models, is still below that of original HR-pQCT scans.  However, this could be overcome by using a hierarchical method that, at least for the high-resolution stages, generates only a subset of slices instead of the entire image.\\
	
	Regarding the applications for research, the ability to synthetically generate realistic, parameterized medical images from a comparatively small set of originals has great potential for enabling algorithmic research. The example at hand is particularly useful to illustrate the possible advantages: as already stated in the introduction, HR-pQCT has well documented advantages over current gold-standard diagnostic bone imaging modalities (i.e. DXA) with regard to the resolution and information to be gained from the imaging. Due to current usage being limited to research applications, obtaining sizeable cohorts of images with a distribution which reflects the average population, especially in younger age groups, can be challenging. However, such cohorts are invaluable for the assessment of potential algorithms for diagnostic and processing applications. While the use of fully synthetic data sets for algorithm training may pose other risks, there are multiple scenarios where augmentation of sample volumes with generated data can be a great advantage. For instance, the ability to customize image attributes may be used to synthesize an optimally distributed test set. The mixing of style-based properties, on the other hand, may be used as a novel form of data augmentation for small data sets, with the ability to generate unique images which show a much larger variance than would be possible with conventional (affine) augmentation techniques. As the ability to rapidly implement graphical user interfaces enables for easy adoption by non-expert users, the number of novel uses for image generation techniques can be expected to rise exponentially in the future.

	\section*{Funding}
	The contribution of C. A. and J.B.P is supported by VASCage – Centre on Clinical Stroke Research. VASCage is a COMET Centre within the Competence Centers for Excellent
	Technologies (COMET) programme and funded by the Federal Ministry for Climate
	Action, Environment, Energy, Mobility, Innovation and Technology, the Federal Ministry of Labour and Economy, and the federal states of Tyrol, Salzburg and Vienna.
	COMET is managed by the Austrian Research Promotion Agency (\"Osterreichische Forschungsf\"orderungsgesellschaft).

	\bibliographystyle{elsarticle-num}
	
	\bibliography{refs.bib}
	\newpage
	
	\appendix

	\section{Generator Configurations}
	\label{app:arch}
	\begin{table}[thb!]
		\begin{minipage}{.425\columnwidth}
			\scriptsize
			\centering
			\caption{Generator details for\textbf{ 3D-ProGAN} with method channel size equal 16. For each layer, information on output channel size (o.c.s), input layer (in) and corresponding activation function (activ.) is provided. Each convolution layer except \textit{out} is followed by a pixelwise feature normalization \cite{karras2017}.}
			\begin{tabularx}{1.01\columnwidth}{|l | l | l | l| l| }
				\toprule
				\textbf{} &\textbf{type} &\textbf{o.c.s.}  &\textbf{in} &\textbf{activ.} \\ \midrule
				d1 & Dense & 8064 & $z$ & \\ \midrule
				r1 & Reshape & 128 &d1& \\ \midrule
				u1 & Upsample & 128 &r1& \\ \midrule
				c11 &Conv\numproduct{3 x 3 x 3}&128&u1&swish \\ \midrule
				c12 &Conv\numproduct{3 x 3 x 3}&128&c11&swish \\ \midrule
				u2 & Upsample & 128 &c12& \\ \midrule
				c21 &Conv\numproduct{3 x 3 x 3}&128&u2&swish \\ \midrule
				c22 &Conv\numproduct{3 x 3 x 3}&128&c21&swish \\ \midrule
				u3 & Upsample & 128 &c22& \\ \midrule
				c31 &Conv\numproduct{3 x 3 x 3}&64&u3&swish \\ \midrule
				c32 &Conv\numproduct{3 x 3 x 3}&64&c31&swish \\ \midrule
				u4 & Upsample & 128 &c32& \\ \midrule
				c41 &Conv\numproduct{3 x 3 x 3}&32&u4&swish \\ \midrule
				c42 &Conv\numproduct{3 x 3 x 3}&32&c41&swish \\ \midrule
				u5 & Upsample & 128 &c42& \\ \midrule
				c51 &Conv\numproduct{3 x 3 x 3}&16&u5&swish \\ \midrule
				c52 &Conv\numproduct{3 x 3 x 3}&16&c51&swish \\ \midrule
				out &Conv\numproduct{3 x 3 x 3}&1&c52&sigmoid \\ \bottomrule
			\end{tabularx}
			\label{tab:3dprogan}
			\vspace{7.2cm}
		\end{minipage}\hfill
		\begin{minipage}{.49\columnwidth}
			\scriptsize
			\centering
			\caption{Generator details for\textbf{ 3D-StyleGAN} with method channel size equal 16. For each layer, information on output channel size (o.c.s), input layer (in) and corresponding activation function (activ.) is provided. The input layer \textit{cc} denotes a constant layer, where the scale is a learned parameter.}
			
			\begin{tabularx}{.999\columnwidth}{|l | l | l | l| l| }
				\toprule
				\textbf{} &\textbf{type} &\textbf{o.c.s.}  &\textbf{in} &\textbf{activ.} \\ \midrule
				m1 & Dense & 512 & $z$ &LReLU \\ \midrule
				m2 & Dense & 512 & m1& LReLU\\ \midrule
				\vdots & \vdots & \vdots &\vdots&\vdots \\ \midrule
				$w$ & Dense & 128 &m5&LReLU \\ \midrule
				s1 & Dense & 128 & $w$ & \\ \midrule
				s2 & Dense & 128 & $w$ & \\ \midrule
				\vdots & \vdots & \vdots &\vdots& \\ \midrule
				s15 & Dense & 16 & $w$ & \\ \midrule

				c11 &Demod\numproduct{3x3x3}&128&cc, s1& \\ \midrule
				c12 & Noise &128 &c11&LReLU\\ \midrule
				up1 & Upsample &128 & c12 & \\ \midrule
				c13 &Demod\numproduct{3x3x3}&128&up1,s2& \\ \midrule
				c14 & Noise &128 &c13&LReLU\\ \midrule
				c15 &Demod\numproduct{3x3x3}&128&c14,s3& \\ \midrule
				c16 & Noise &128 &c15&LReLU\\ \midrule
				
				c21 &Demod\numproduct{3x3x3}&128&c16,s4& \\ \midrule
				c22 & Noise &128 &c21&LReLU\\ \midrule
				up2 & Upsample &128 & c22 & \\ \midrule
				c23 &Demod\numproduct{3x3x3}&128&up2,s5& \\ \midrule
				c24 & Noise &128 &c23&LReLU\\ \midrule
				c25 &Demod\numproduct{3x3x3}&128&c24,s6& \\ \midrule
				c26 & Noise &128 &c25&LReLU\\ \midrule
				
				\vdots & \vdots & \vdots &\vdots&\vdots \\ \midrule
				
				c51 &Demod\numproduct{3x3x3}&16&c46, s13& \\ \midrule
				c52 & Noise &16 &c51&LReLU\\ \midrule
				up5 & Upsample &16 & c52 & \\ \midrule
				c53 &Demod\numproduct{3x3x3}&128&up5,s14& \\ \midrule
				c54 & Noise &16 &c53&LReLU\\ \midrule
				c55 &Demod\numproduct{3x3x3}&128&c54,s15& \\ \midrule
				c56 & Noise &16 &c55&LReLU\\ \midrule
				
				out &Conv\numproduct{3x3x3}&1&c56&sigmoid \\ \bottomrule
			\end{tabularx}
			\label{tab:3dstylegan}
		\end{minipage}
		
	\end{table}

	\section{Training Details}
	\label{app:train}
	
	The generators in 3D-ProGAN and 3D-StyleGAN have been trained using Adam optimizer with hyper-parameters $\beta_1 = 0, \beta_2=0.98, \epsilon=1e-7$ and different learning rates $\alpha \in \{k\times 10^{-3}\mid k=2,\ldots,6\}$. Gradient norm clipping with threshold 2 is applied at each step. The concept of equalized learning rates are used, thus all convolution and dense layers are initialized using standard normal distribution. For 3D-StyleGAN, the learning rates of the latent space mapping network $\Phi$ are multiplied by factor $0.02$.\\
	
	The critic networks have been trained using Adam optimizer with hyper-parameters $\beta_1 = 0,\beta_2=0.98,\epsilon=5e-5$ and different learning rates $\alpha \in \{k\times 10^{-3}\mid k=2,\ldots,6\}$. One generator update is followed by $n_c$ critic updates, where $n_c \in \{5,6,7,8\}$ during the experiments.
	Training at stage 1 is continued until the critic has seen 180k samples. Training on stage 2 is conducted for 360k scans, while transition of the new layers takes place for the first 180k samples. The procedure is continued until the model reaches final stage 5. Every time a new stage is reached, the learning rates are multiplied by factor 0.85. The size of the minibatches for stage 1 to 5 equals $24,24,12,6,3$. Training lasts approximately three days on a single NVIDIA A100 40GB GPU.
	\vspace{1cm}
	
	\section{Further Visualizations}
	In the following, more synthetic HR-pQCT instances sampled from both proposed methods, 3D-ProGAN and 3D-StyleGAN, are visualized. Each column represents a different level of truncation. For both generation methods, greater use of the truncation trick increases image quality, but at the cost of reduced synthesis diversity.
	
	\begin{figure}[thb!]
		\includegraphics[height=.11\textheight]{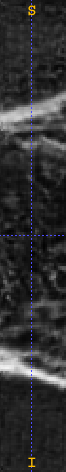}\hspace{-.05cm}
		\includegraphics[height=.11\textheight]{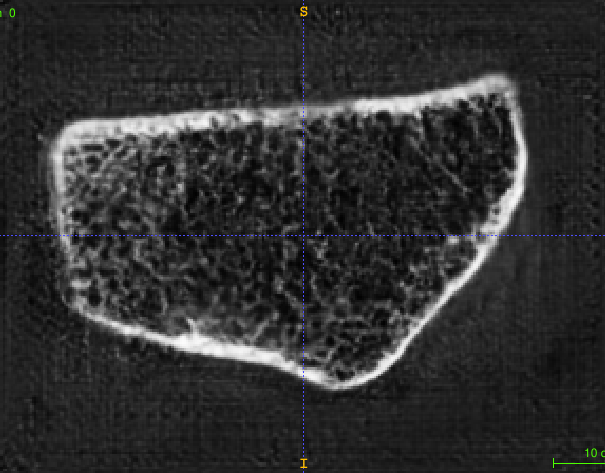}\hspace{.15cm}
		\includegraphics[height=.11\textheight]{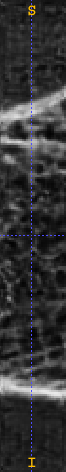}\hspace{-.05cm}
		\includegraphics[height=.11\textheight]{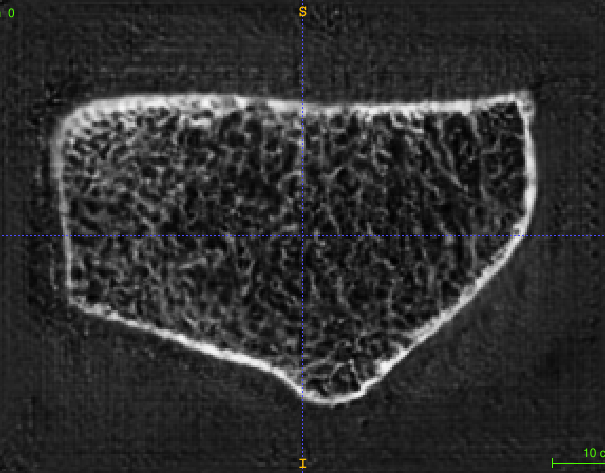}\hspace{.15cm}
		\includegraphics[height=.11\textheight]{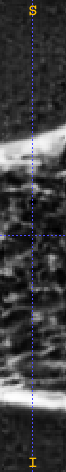}\hspace{-.05cm}
		\includegraphics[height=.11\textheight]{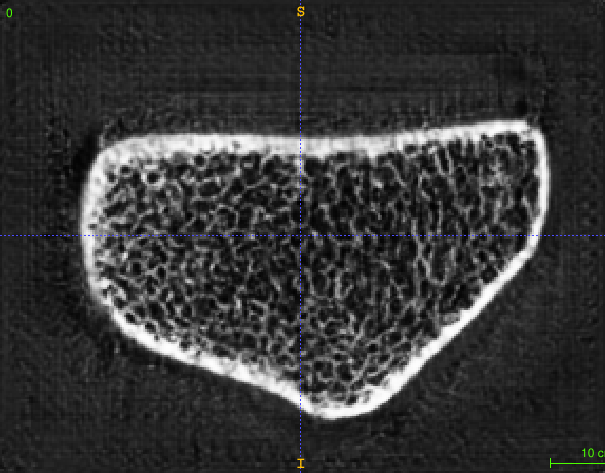}\hspace{.15cm}
		\includegraphics[height=.11\textheight]{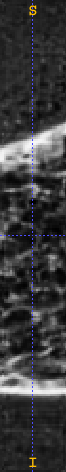}\hspace{-.05cm}
		\includegraphics[height=.11\textheight]{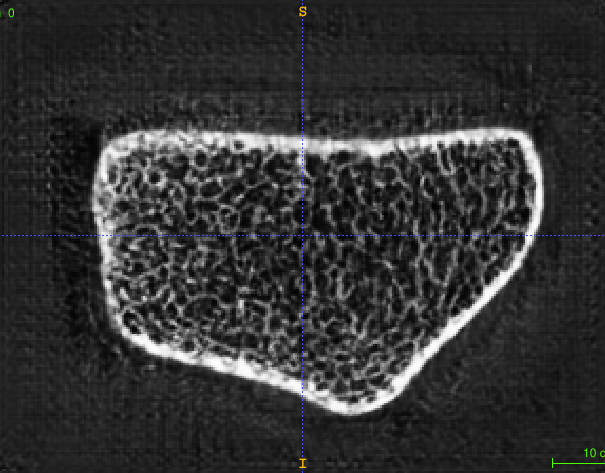}\vspace{-.06cm}
		
		\hspace{.25cm}
		\includegraphics[width=.197\textwidth]{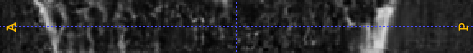}\hspace{.59cm}
		\includegraphics[width=.197\textwidth]{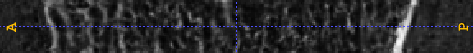}\hspace{.59cm}
		\includegraphics[width=.197\textwidth]{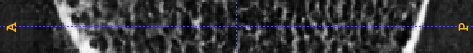}\hspace{.59cm}
		\includegraphics[width=.197\textwidth]{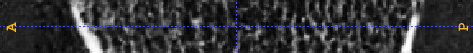}
		\vspace{.2cm}
		
		\includegraphics[height=.11\textheight]{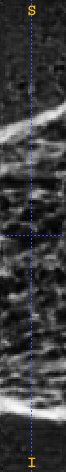}\hspace{-.05cm}
		\includegraphics[height=.11\textheight]{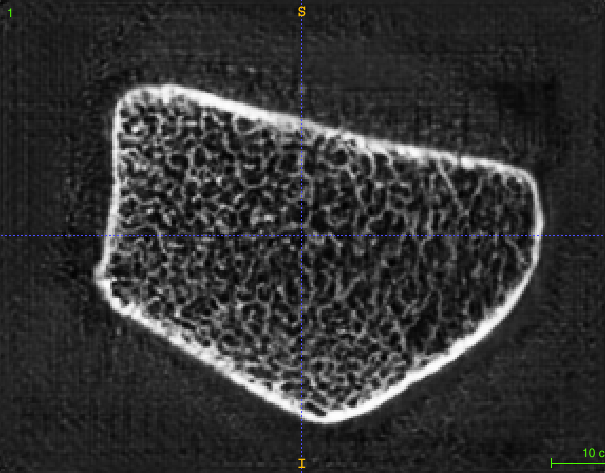}\hspace{.15cm}
		\includegraphics[height=.11\textheight]{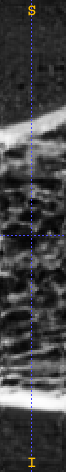}\hspace{-.05cm}
		\includegraphics[height=.11\textheight]{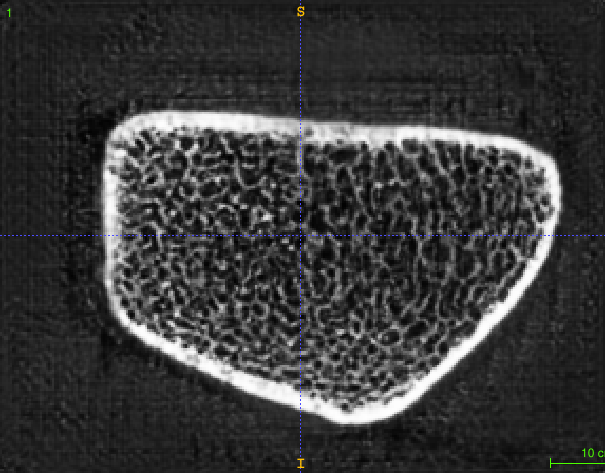}\hspace{.15cm}
		\includegraphics[height=.11\textheight]{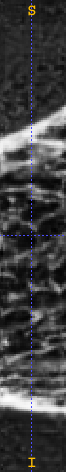}\hspace{-.05cm}
		\includegraphics[height=.11\textheight]{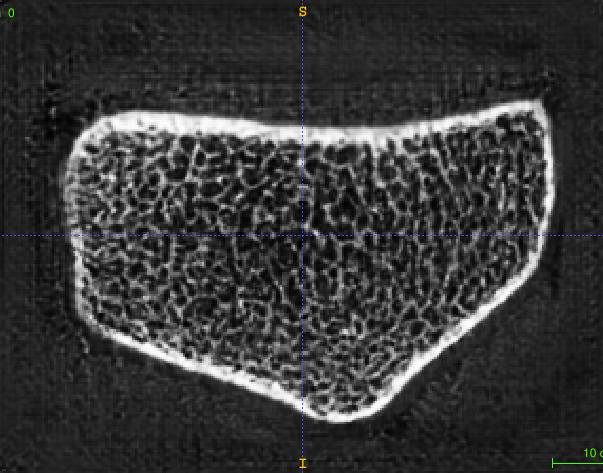}\hspace{.15cm}
		\includegraphics[height=.11\textheight]{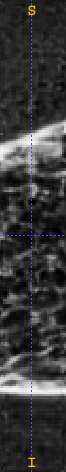}\hspace{-.05cm}
		\includegraphics[height=.11\textheight]{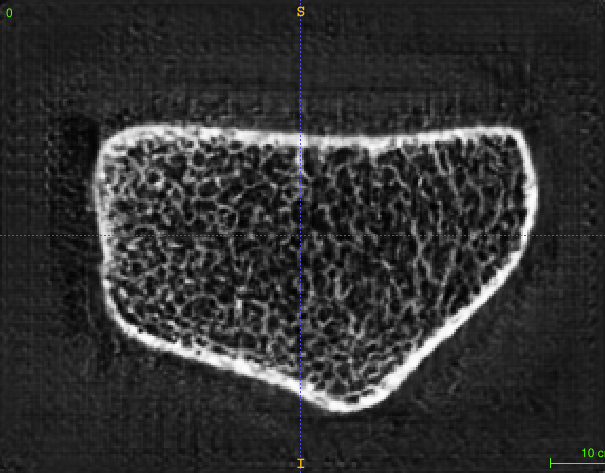}\vspace{-.06cm}
		
		\hspace{.26cm}
		\includegraphics[width=.197\textwidth]{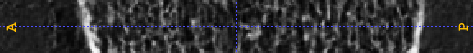}\hspace{.59cm}
		\includegraphics[width=.197\textwidth]{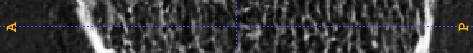}\hspace{.59cm}
		\includegraphics[width=.197\textwidth]{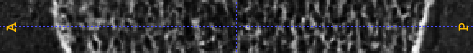}\hspace{.59cm}
		\includegraphics[width=.197\textwidth]{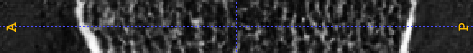}
		\vspace{.2cm}
		
		\includegraphics[height=.11\textheight]{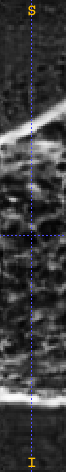}\hspace{-.05cm}
		\includegraphics[height=.11\textheight]{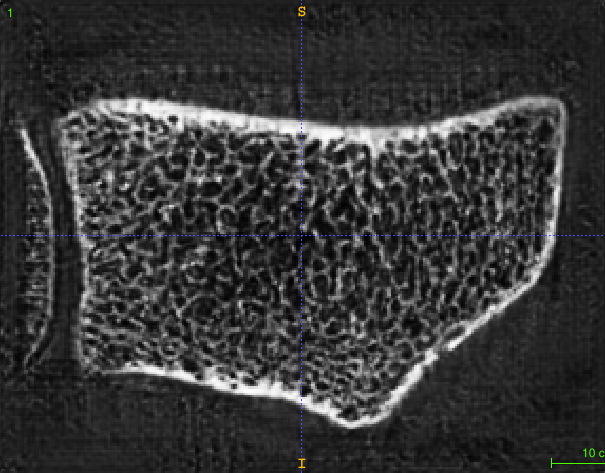}\hspace{.15cm}
		\includegraphics[height=.11\textheight]{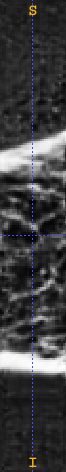}\hspace{-.05cm}
		\includegraphics[height=.11\textheight]{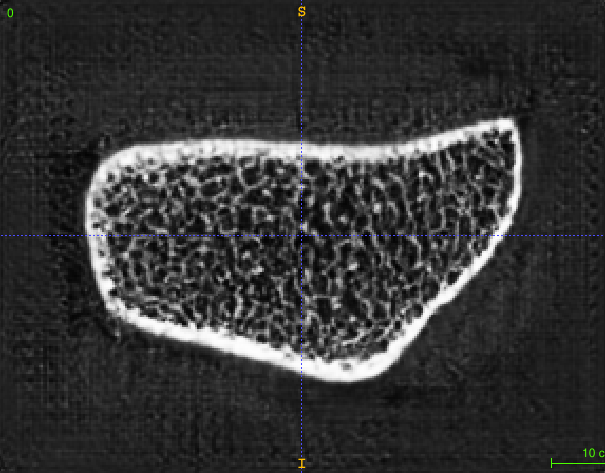}\hspace{.15cm}
		\includegraphics[height=.11\textheight]{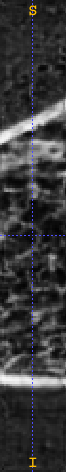}\hspace{-.05cm}
		\includegraphics[height=.11\textheight]{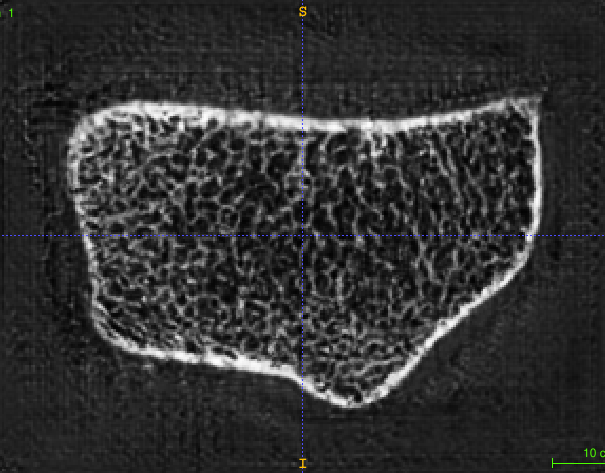}\hspace{.15cm}
		\includegraphics[height=.11\textheight]{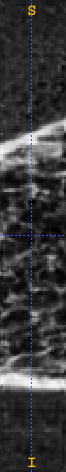}\hspace{-.05cm}
		\includegraphics[height=.11\textheight]{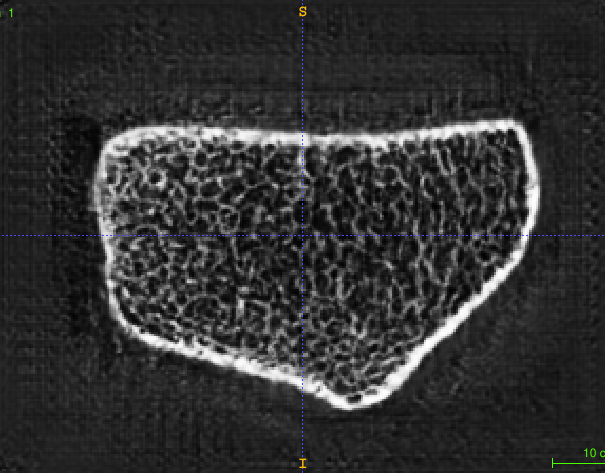}\vspace{-.06cm}
		
		\hspace{.26cm}
		\includegraphics[width=.197\textwidth]{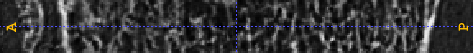}\hspace{.59cm}
		\includegraphics[width=.197\textwidth]{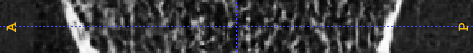}\hspace{.59cm}
		\includegraphics[width=.197\textwidth]{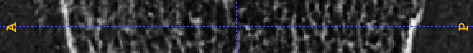}\hspace{.59cm}
		\includegraphics[width=.197\textwidth]{figs_arxiv/pggan_trunc/0.2/02_0.png}
		\vspace{.2cm}
		
		\includegraphics[height=.11\textheight]{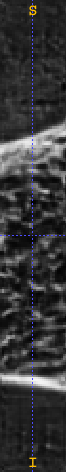}\hspace{-.05cm}
		\includegraphics[height=.11\textheight]{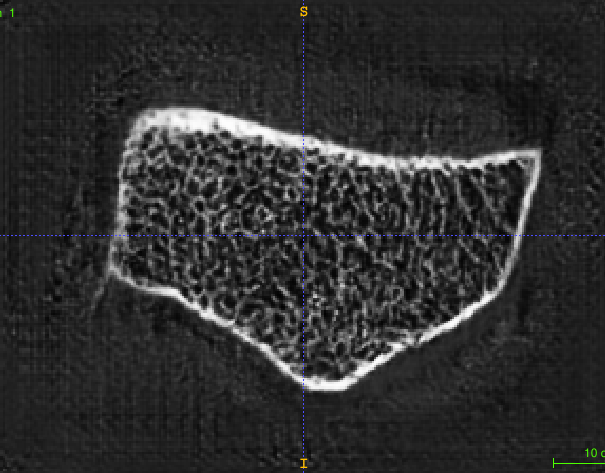}\hspace{.15cm}
		\includegraphics[height=.11\textheight]{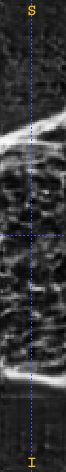}\hspace{-.05cm}
		\includegraphics[height=.11\textheight]{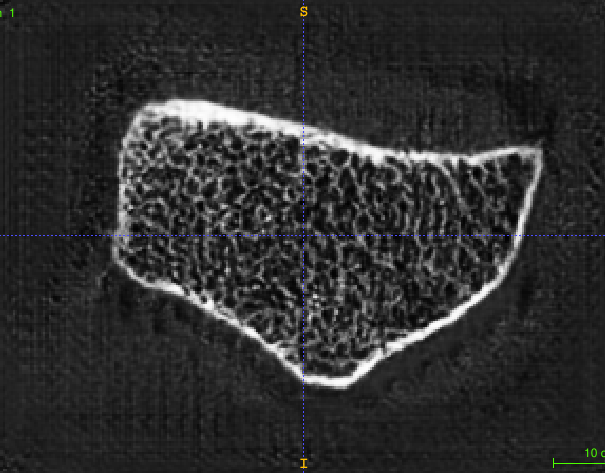}\hspace{.15cm}
		\includegraphics[height=.11\textheight]{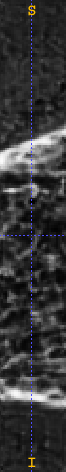}\hspace{-.05cm}
		\includegraphics[height=.11\textheight]{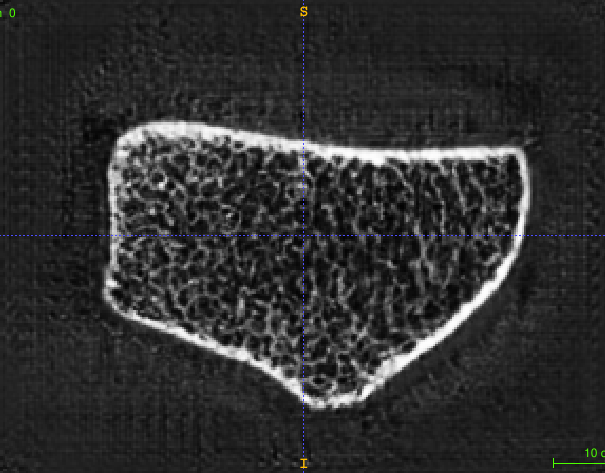}\hspace{.15cm}
		\includegraphics[height=.11\textheight]{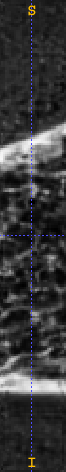}\hspace{-.05cm}
		\includegraphics[height=.11\textheight]{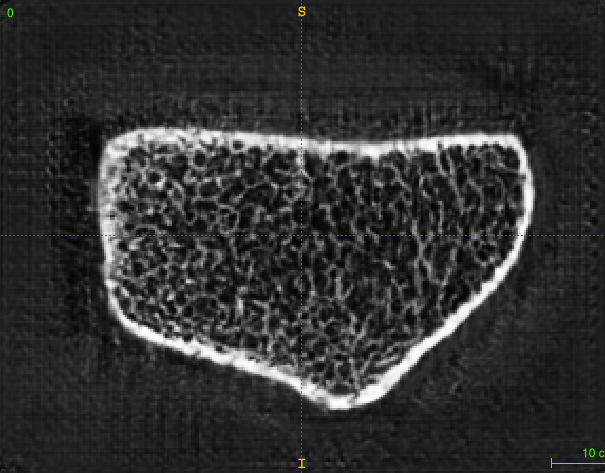}\vspace{-.06cm}
		
		\hspace{.26cm}
		\includegraphics[width=.197\textwidth]{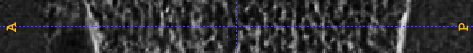}\hspace{.59cm}
		\includegraphics[width=.197\textwidth]{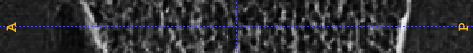}\hspace{.59cm}
		\includegraphics[width=.197\textwidth]{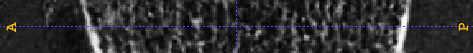}\hspace{.59cm}
		\includegraphics[width=.197\textwidth]{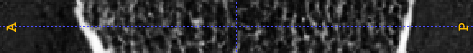}
		\vspace{.2cm}
		
		\includegraphics[height=.11\textheight]{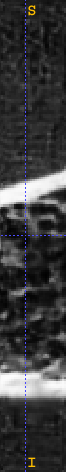}\hspace{-.05cm}
		\includegraphics[height=.11\textheight]{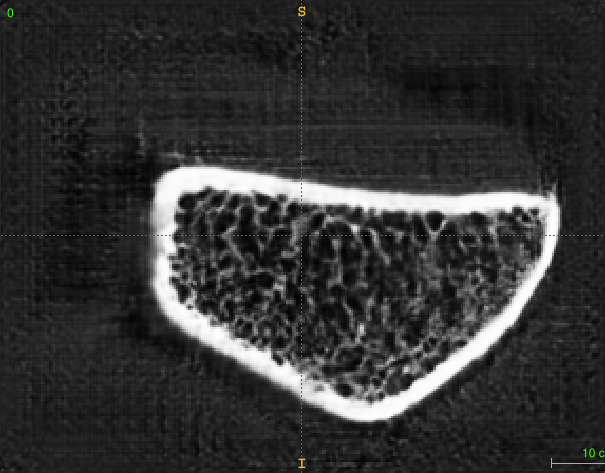}\hspace{.15cm}
		\includegraphics[height=.11\textheight]{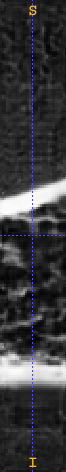}\hspace{-.05cm}
		\includegraphics[height=.11\textheight]{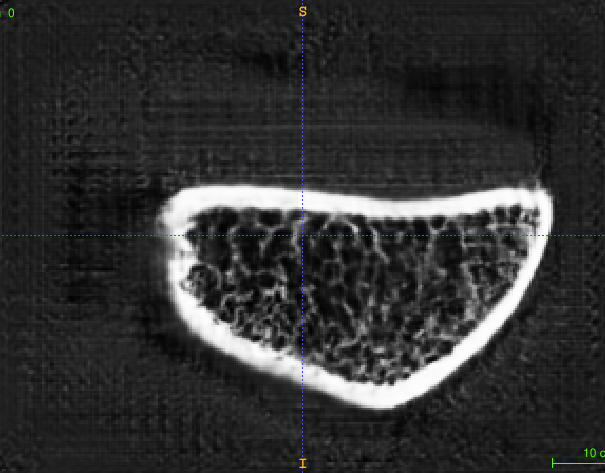}\hspace{.15cm}
		\includegraphics[height=.11\textheight]{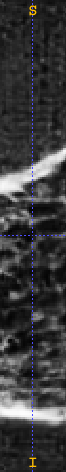}\hspace{-.05cm}
		\includegraphics[height=.11\textheight]{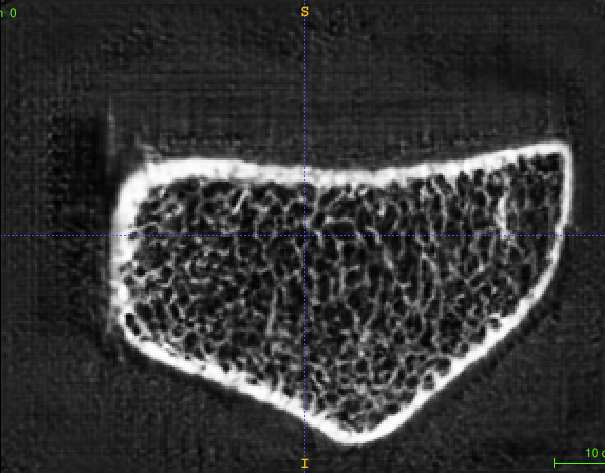}\hspace{.15cm}
		\includegraphics[height=.11\textheight]{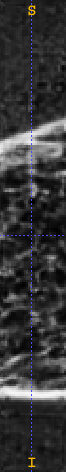}\hspace{-.05cm}
		\includegraphics[height=.11\textheight]{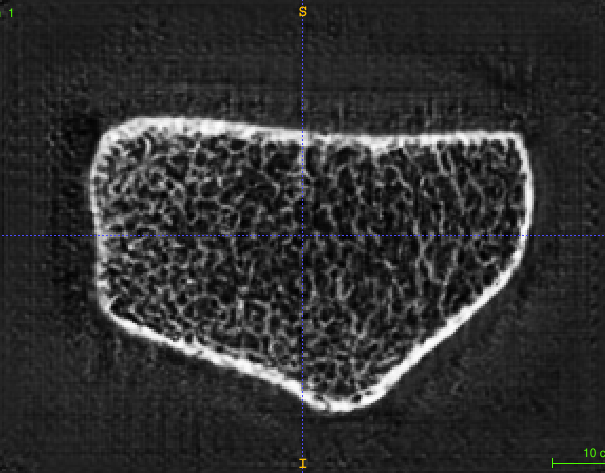}\vspace{-.06cm}
		
		\hspace{.26cm}
		\includegraphics[width=.197\textwidth]{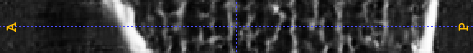}\hspace{.59cm}
		\includegraphics[width=.197\textwidth]{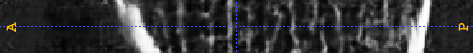}\hspace{.59cm}
		\includegraphics[width=.197\textwidth]{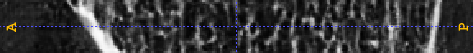}\hspace{.59cm}
		\includegraphics[width=.197\textwidth]{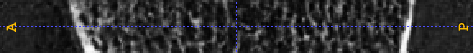}
		\vspace{.2cm}
		
		\includegraphics[height=.11\textheight]{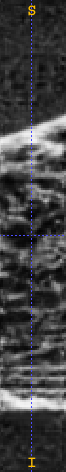}\hspace{-.05cm}
		\includegraphics[height=.11\textheight]{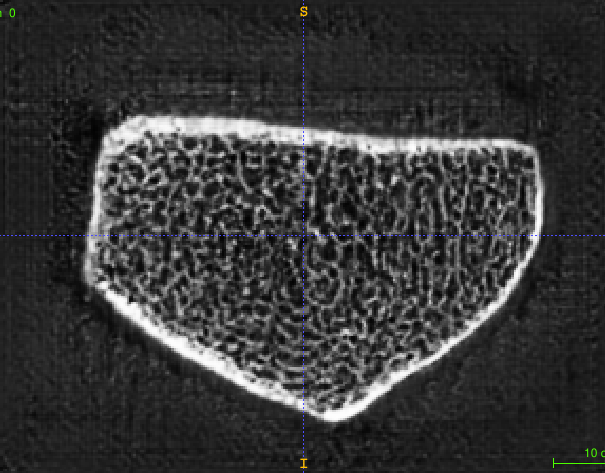}\hspace{.15cm}
		\includegraphics[height=.11\textheight]{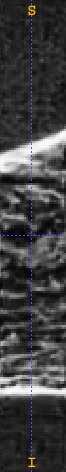}\hspace{-.05cm}
		\includegraphics[height=.11\textheight]{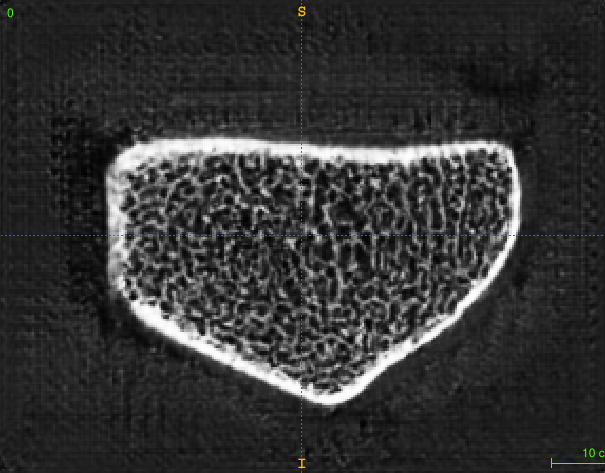}\hspace{.15cm}
		\includegraphics[height=.11\textheight]{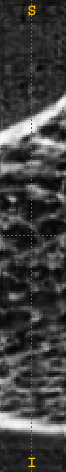}\hspace{-.05cm}
		\includegraphics[height=.11\textheight]{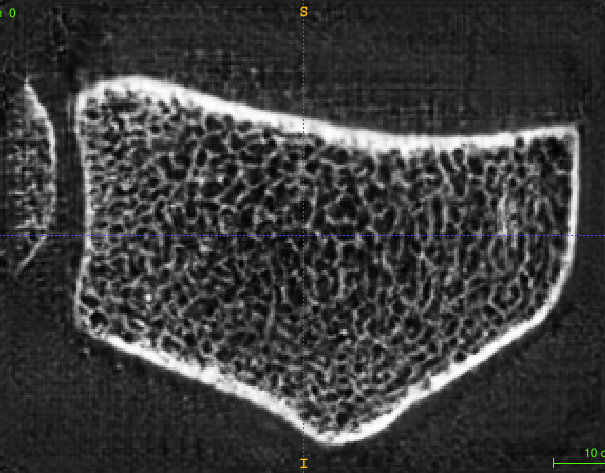}\hspace{.15cm}
		\includegraphics[height=.11\textheight]{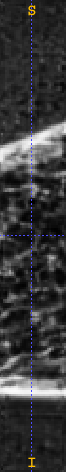}\hspace{-.05cm}
		\includegraphics[height=.11\textheight]{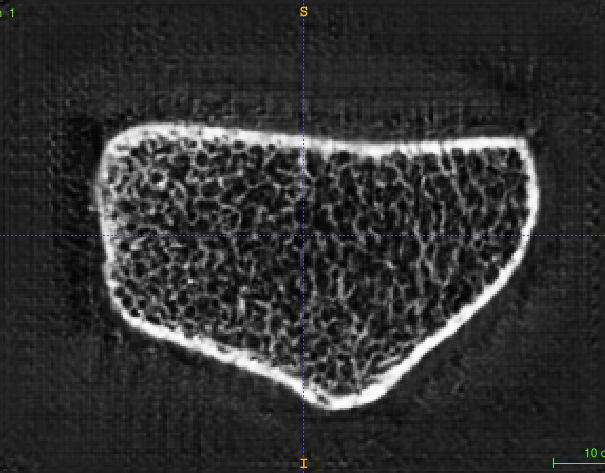}\vspace{-.06cm}
		
		\hspace{.26cm}
		\includegraphics[width=.197\textwidth]{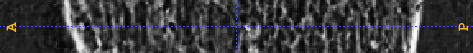}\hspace{.59cm}
		\includegraphics[width=.197\textwidth]{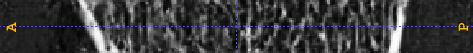}\hspace{.59cm}
		\includegraphics[width=.197\textwidth]{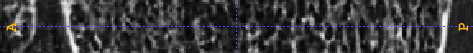}\hspace{.59cm}
		\includegraphics[width=.197\textwidth]{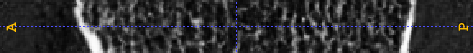}
		\vspace{.2cm}
		
		\includegraphics[height=.11\textheight]{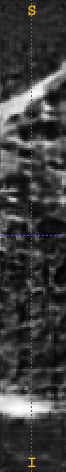}\hspace{-.05cm}
		\includegraphics[height=.11\textheight]{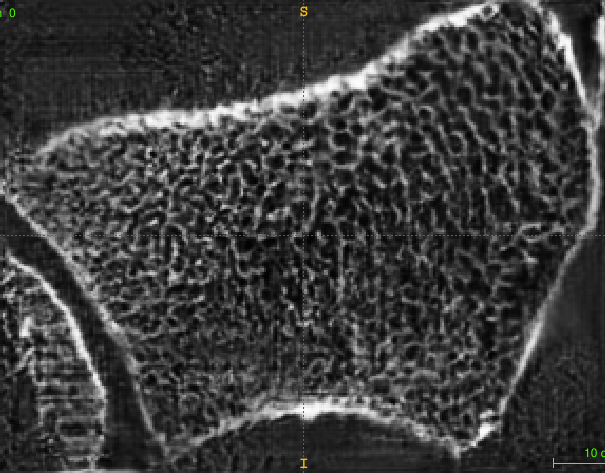}\hspace{.15cm}
		\includegraphics[height=.11\textheight]{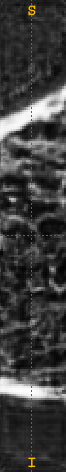}\hspace{-.05cm}
		\includegraphics[height=.11\textheight]{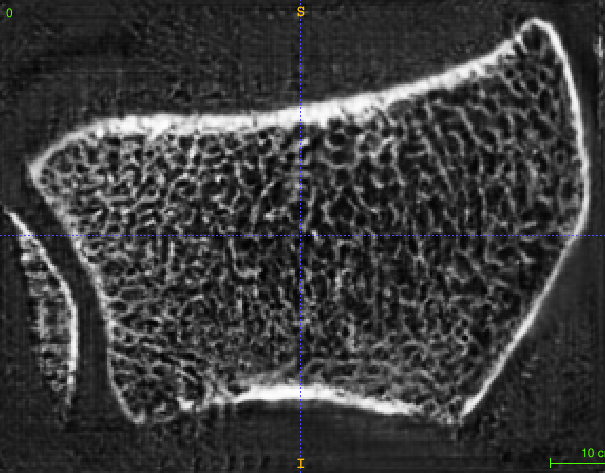}\hspace{.15cm}
		\includegraphics[height=.11\textheight]{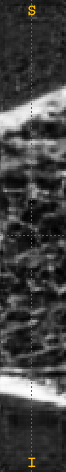}\hspace{-.05cm}
		\includegraphics[height=.11\textheight]{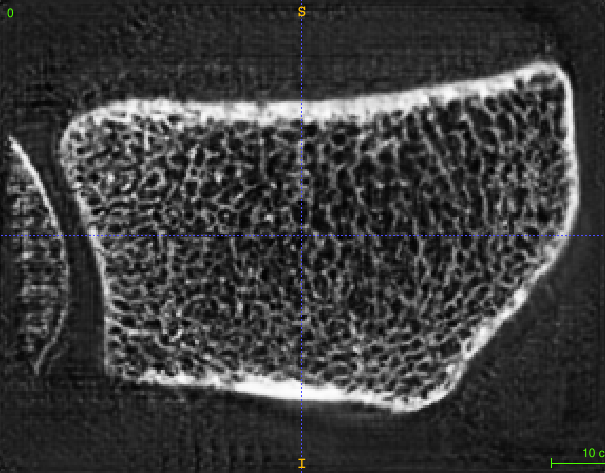}\hspace{.15cm}
		\includegraphics[height=.11\textheight]{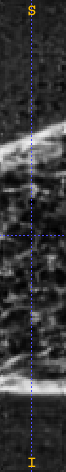}\hspace{-.05cm}
		\includegraphics[height=.11\textheight]{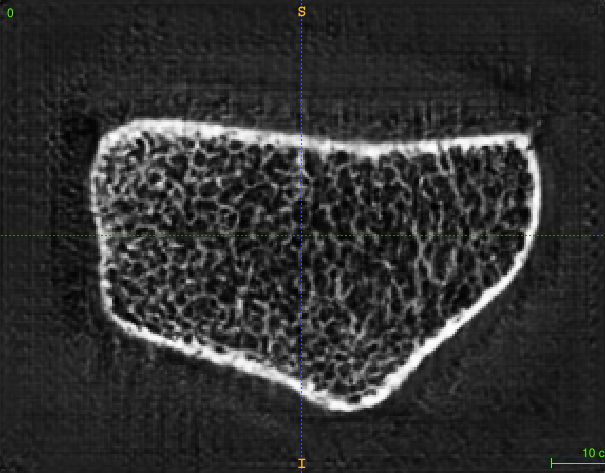}\vspace{-.06cm}
		
		\hspace{.26cm}
		\includegraphics[width=.197\textwidth]{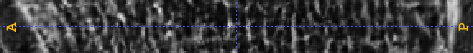}\hspace{.59cm}
		\includegraphics[width=.197\textwidth]{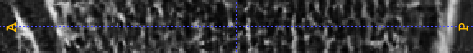}\hspace{.59cm}
		\includegraphics[width=.197\textwidth]{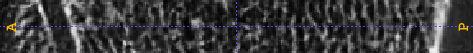}\hspace{.59cm}
		\includegraphics[width=.197\textwidth]{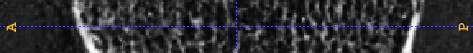}
		
		\caption{Synthetic HR-pQCT volumes sampled from the proposed 3D-ProGAN approach with varying parameter for the truncated normal distribution. From left to right column: truncation parameter equals $\{2.6,1.8,1,0.2\}$.}
		\label{fig:pggan_trunc}
	\end{figure}

	\newpage
	
	\begin{figure}[thb!]
		\includegraphics[height=.11\textheight]{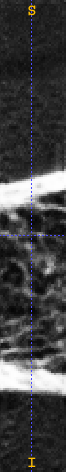}\hspace{-.05cm}
		\includegraphics[height=.11\textheight]{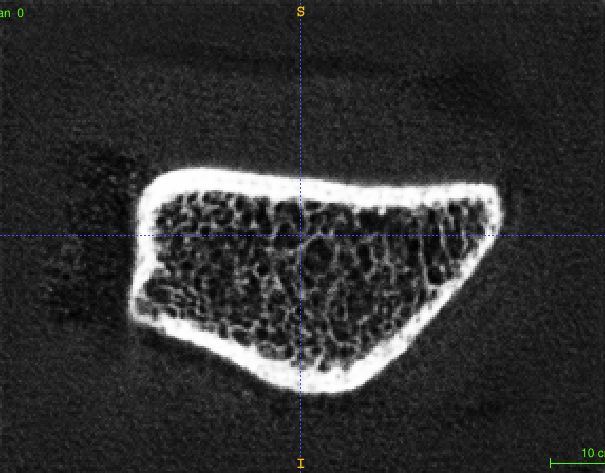}\hspace{.15cm}
		\includegraphics[height=.11\textheight]{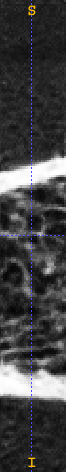}\hspace{-.05cm}
		\includegraphics[height=.11\textheight]{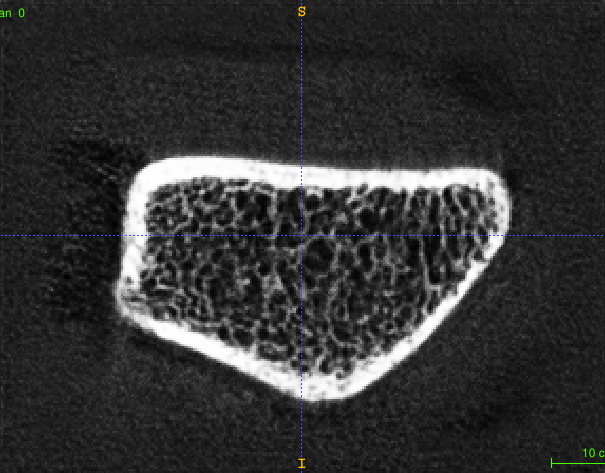}\hspace{.15cm}
		\includegraphics[height=.11\textheight]{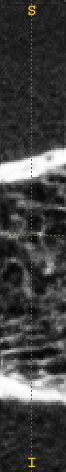}\hspace{-.05cm}
		\includegraphics[height=.11\textheight]{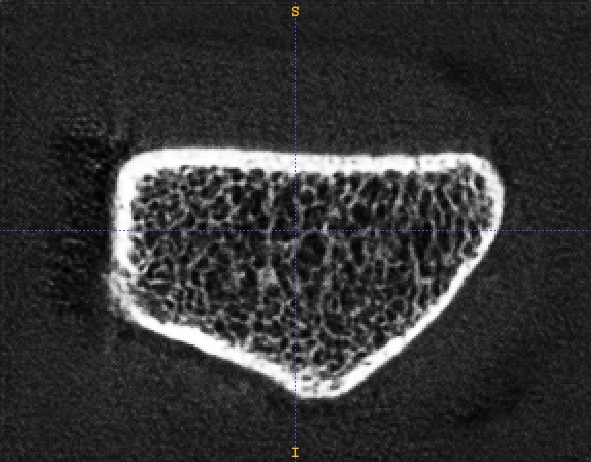}\hspace{.15cm}
		\includegraphics[height=.11\textheight]{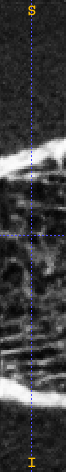}\hspace{-.05cm}
		\includegraphics[height=.11\textheight]{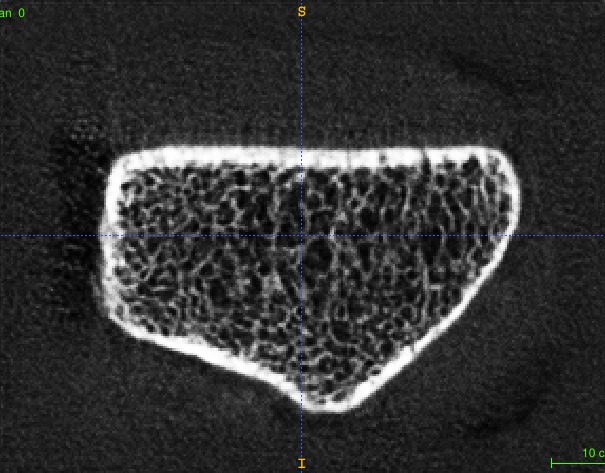}\vspace{-.06cm}
		
		\hspace{.25cm}
		\includegraphics[width=.197\textwidth]{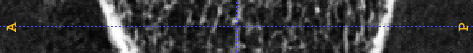}\hspace{.59cm}
		\includegraphics[width=.197\textwidth]{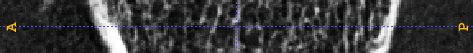}\hspace{.59cm}
		\includegraphics[width=.197\textwidth]{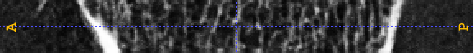}\hspace{.59cm}
		\includegraphics[width=.197\textwidth]{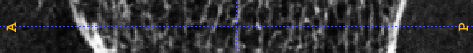}
		\vspace{.2cm}
		
		\includegraphics[height=.11\textheight]{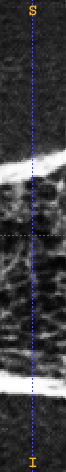}\hspace{-.05cm}
		\includegraphics[height=.11\textheight]{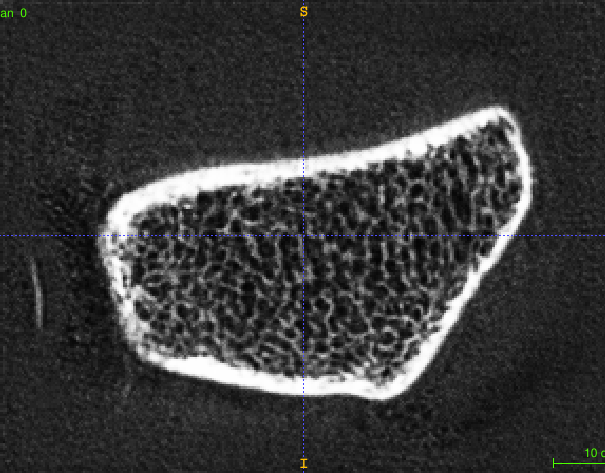}\hspace{.15cm}
		\includegraphics[height=.11\textheight]{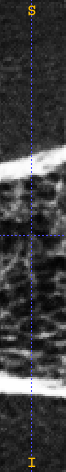}\hspace{-.05cm}
		\includegraphics[height=.11\textheight]{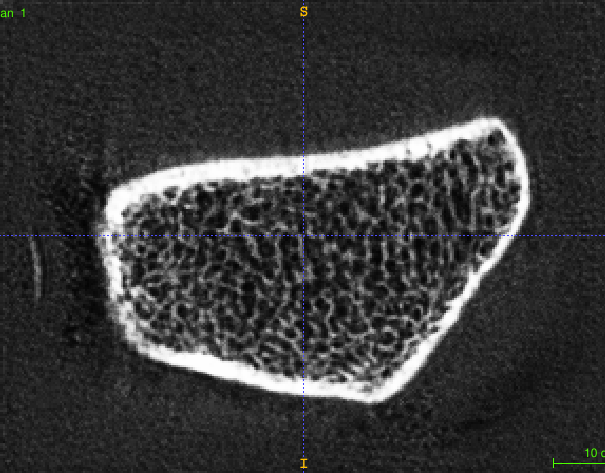}\hspace{.15cm}
		\includegraphics[height=.11\textheight]{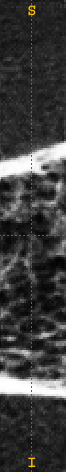}\hspace{-.05cm}
		\includegraphics[height=.11\textheight]{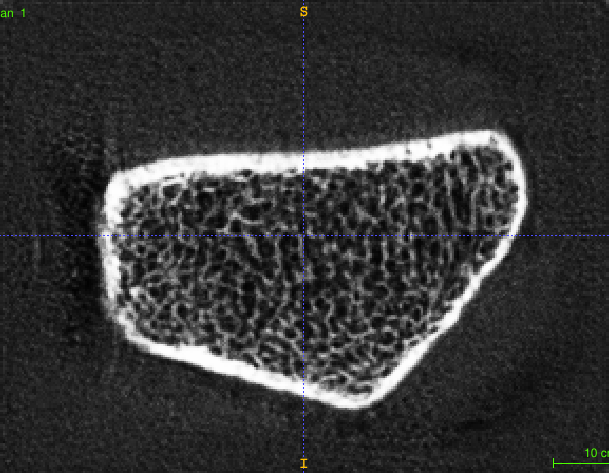}\hspace{.15cm}
		\includegraphics[height=.11\textheight]{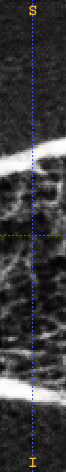}\hspace{-.05cm}
		\includegraphics[height=.11\textheight]{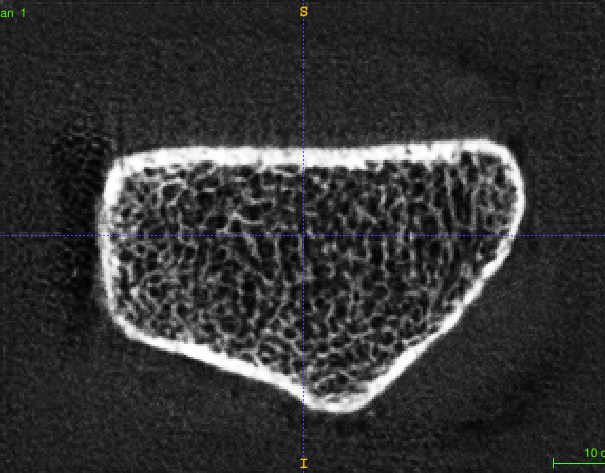}\vspace{-.06cm}
		
		\hspace{.25cm}
		\includegraphics[width=.197\textwidth]{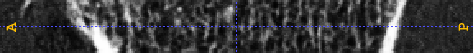}\hspace{.59cm}
		\includegraphics[width=.197\textwidth]{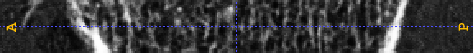}\hspace{.59cm}
		\includegraphics[width=.197\textwidth]{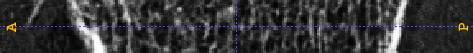}\hspace{.59cm}
		\includegraphics[width=.197\textwidth]{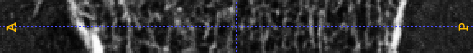}
		\vspace{.2cm}
		
		\includegraphics[height=.11\textheight]{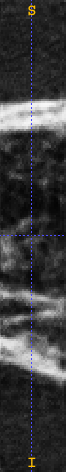}\hspace{-.05cm}
		\includegraphics[height=.11\textheight]{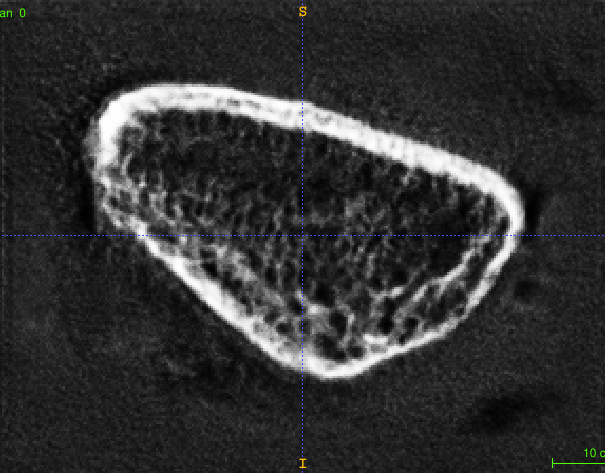}\hspace{.15cm}
		\includegraphics[height=.11\textheight]{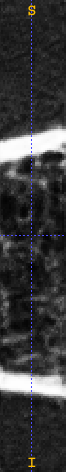}\hspace{-.05cm}
		\includegraphics[height=.11\textheight]{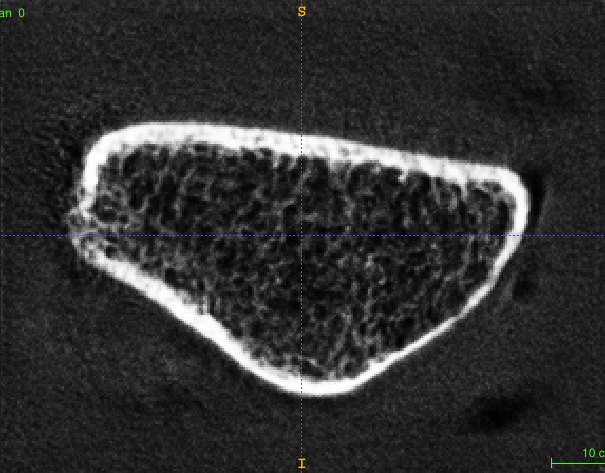}\hspace{.15cm}
		\includegraphics[height=.11\textheight]{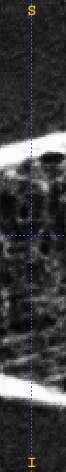}\hspace{-.05cm}
		\includegraphics[height=.11\textheight]{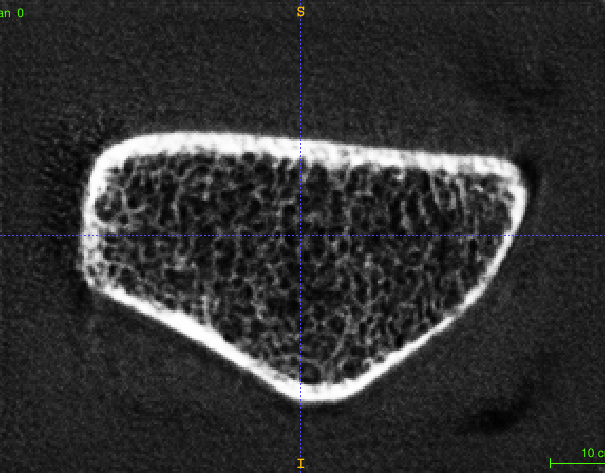}\hspace{.15cm}
		\includegraphics[height=.11\textheight]{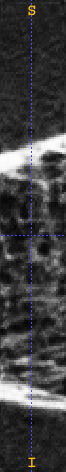}\hspace{-.05cm}
		\includegraphics[height=.11\textheight]{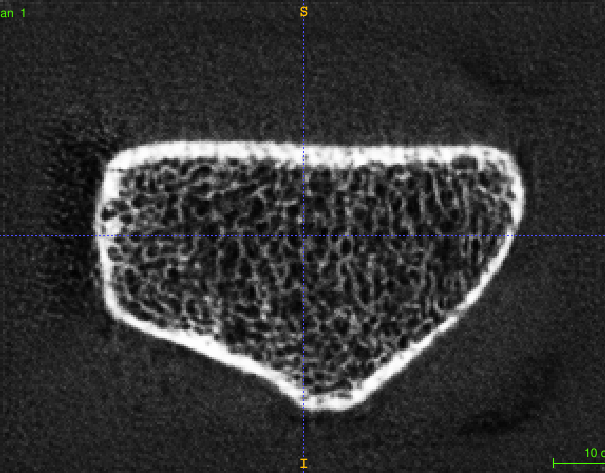}\vspace{-.06cm}
		
		\hspace{.25cm}
		\includegraphics[width=.197\textwidth]{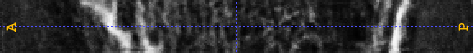}\hspace{.59cm}
		\includegraphics[width=.197\textwidth]{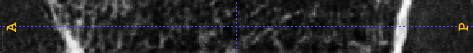}\hspace{.59cm}
		\includegraphics[width=.197\textwidth]{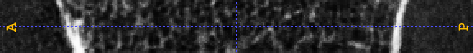}\hspace{.59cm}
		\includegraphics[width=.197\textwidth]{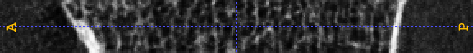}
		\vspace{.2cm}

		\includegraphics[height=.11\textheight]{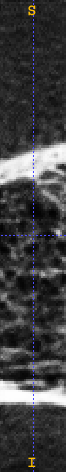}\hspace{-.05cm}
		\includegraphics[height=.11\textheight]{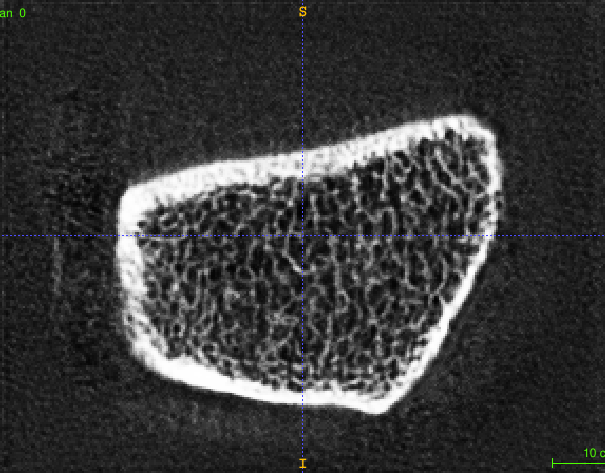}\hspace{.15cm}
		\includegraphics[height=.11\textheight]{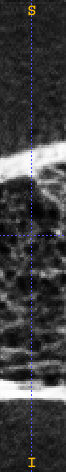}\hspace{-.05cm}
		\includegraphics[height=.11\textheight]{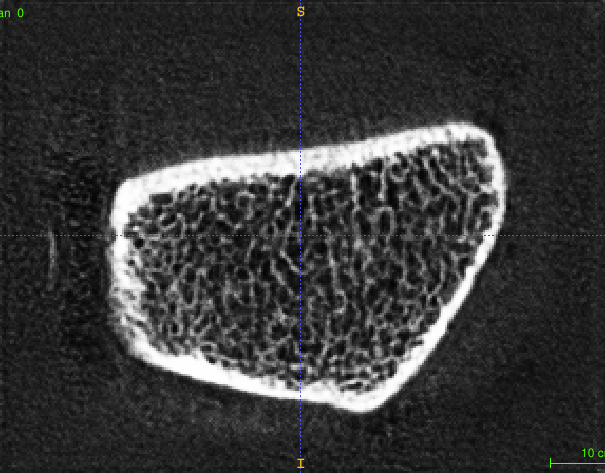}\hspace{.15cm}
		\includegraphics[height=.11\textheight]{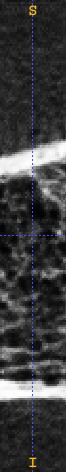}\hspace{-.05cm}
		\includegraphics[height=.11\textheight]{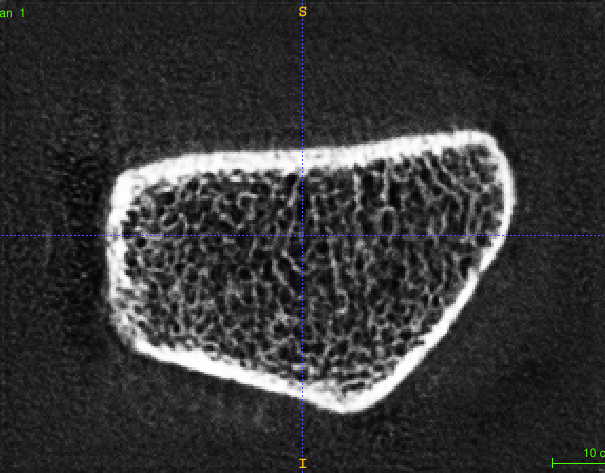}\hspace{.15cm}
		\includegraphics[height=.11\textheight]{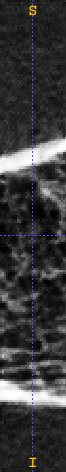}\hspace{-.05cm}
		\includegraphics[height=.11\textheight]{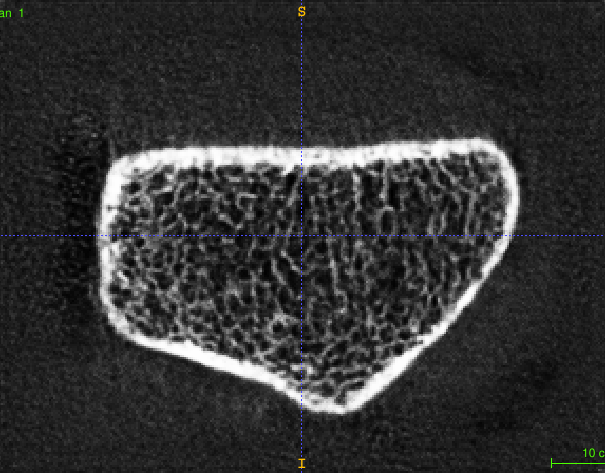}\vspace{-.06cm}
		
		\hspace{.25cm}
		\includegraphics[width=.197\textwidth]{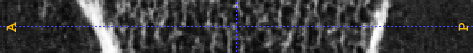}\hspace{.59cm}
		\includegraphics[width=.197\textwidth]{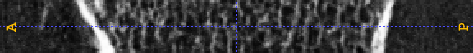}\hspace{.59cm}
		\includegraphics[width=.197\textwidth]{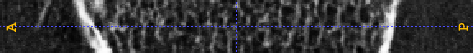}\hspace{.59cm}
		\includegraphics[width=.197\textwidth]{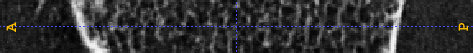}
		\vspace{.2cm}
		
		\includegraphics[height=.11\textheight]{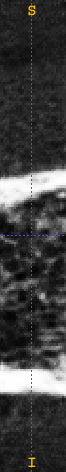}\hspace{-.05cm}
		\includegraphics[height=.11\textheight]{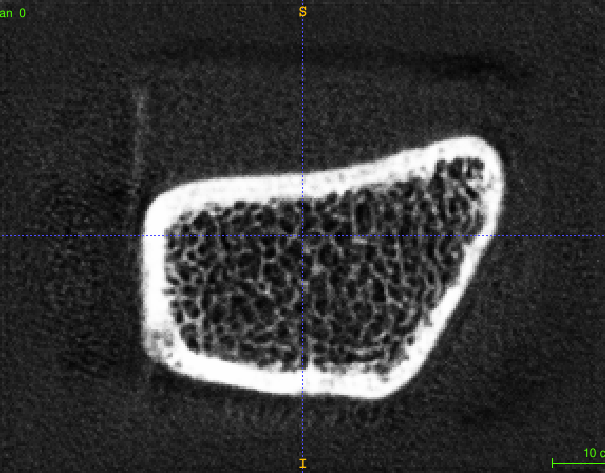}\hspace{.15cm}
		\includegraphics[height=.11\textheight]{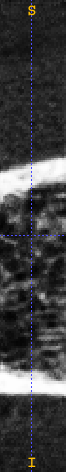}\hspace{-.05cm}
		\includegraphics[height=.11\textheight]{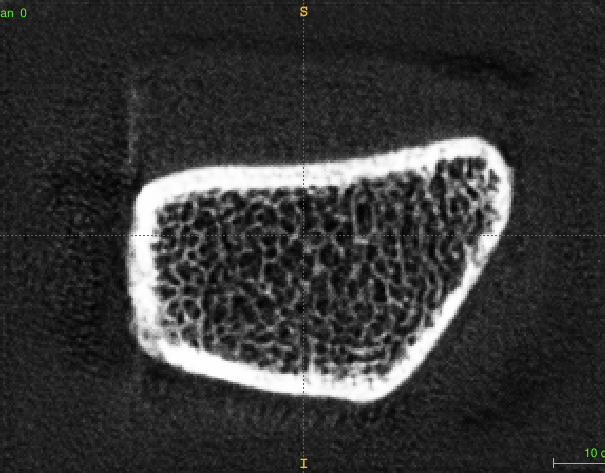}\hspace{.15cm}
		\includegraphics[height=.11\textheight]{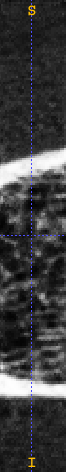}\hspace{-.05cm}
		\includegraphics[height=.11\textheight]{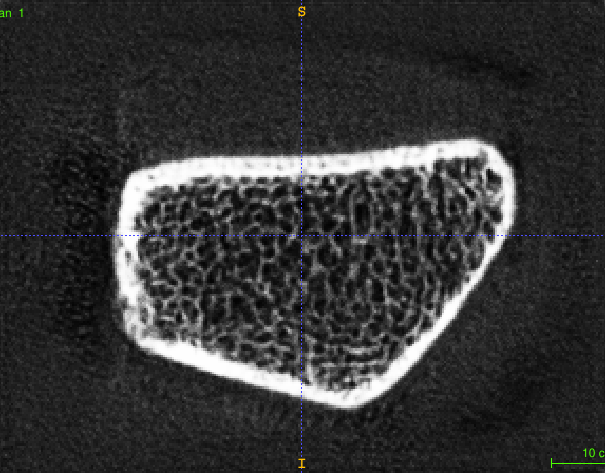}\hspace{.15cm}
		\includegraphics[height=.11\textheight]{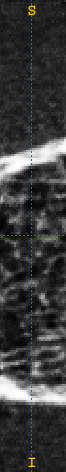}\hspace{-.05cm}
		\includegraphics[height=.11\textheight]{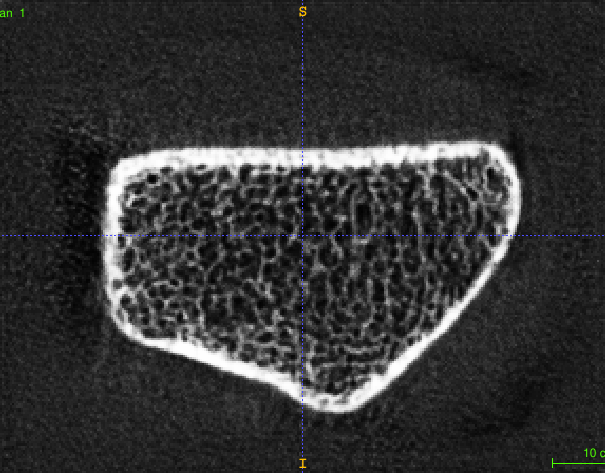}\vspace{-.06cm}
		
		\hspace{.25cm}
		\includegraphics[width=.197\textwidth]{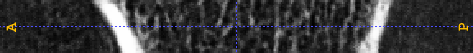}\hspace{.59cm}
		\includegraphics[width=.197\textwidth]{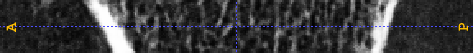}\hspace{.59cm}
		\includegraphics[width=.197\textwidth]{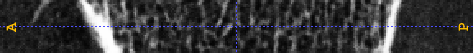}\hspace{.59cm}
		\includegraphics[width=.197\textwidth]{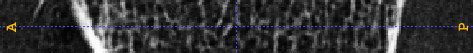}
		\vspace{.2cm}
		
		\includegraphics[height=.11\textheight]{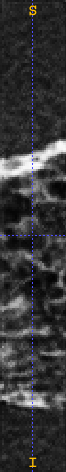}\hspace{-.05cm}
		\includegraphics[height=.11\textheight]{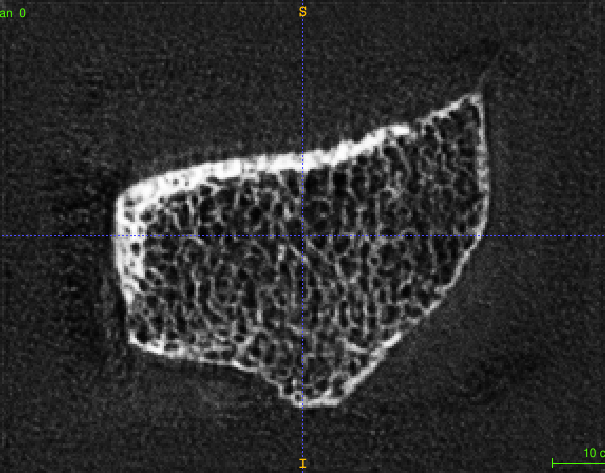}\hspace{.15cm}
		\includegraphics[height=.11\textheight]{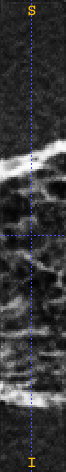}\hspace{-.05cm}
		\includegraphics[height=.11\textheight]{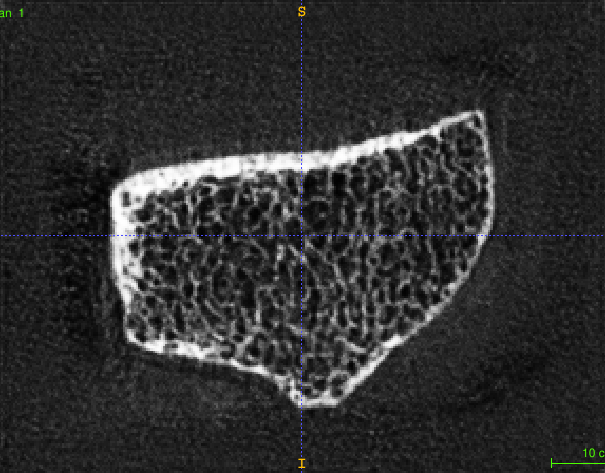}\hspace{.15cm}
		\includegraphics[height=.11\textheight]{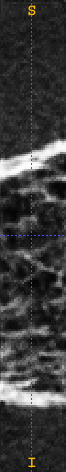}\hspace{-.05cm}
		\includegraphics[height=.11\textheight]{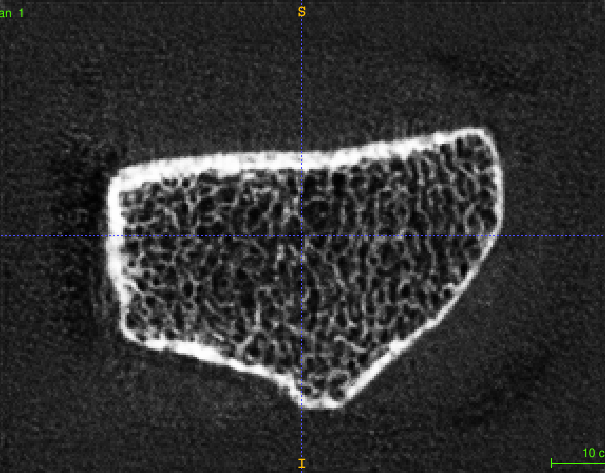}\hspace{.15cm}
		\includegraphics[height=.11\textheight]{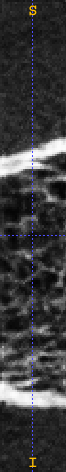}\hspace{-.05cm}
		\includegraphics[height=.11\textheight]{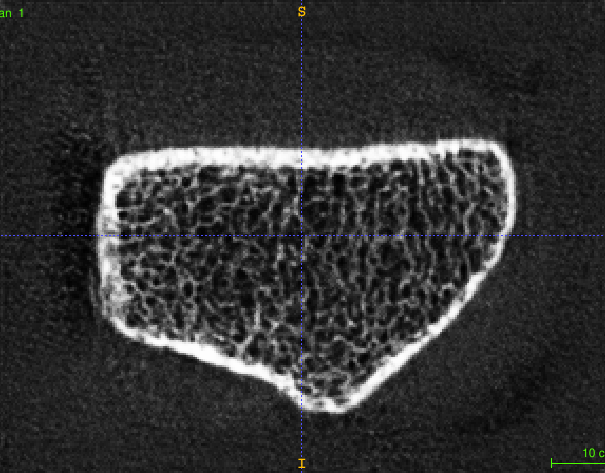}\vspace{-.06cm}
		
		\hspace{.25cm}
		\includegraphics[width=.197\textwidth]{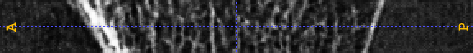}\hspace{.59cm}
		\includegraphics[width=.197\textwidth]{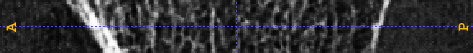}\hspace{.59cm}
		\includegraphics[width=.197\textwidth]{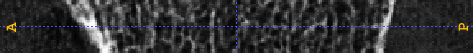}\hspace{.59cm}
		\includegraphics[width=.197\textwidth]{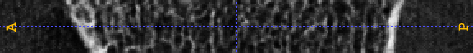}
		\vspace{.2cm}
		
		\includegraphics[height=.11\textheight]{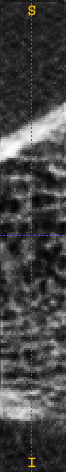}\hspace{-.05cm}
		\includegraphics[height=.11\textheight]{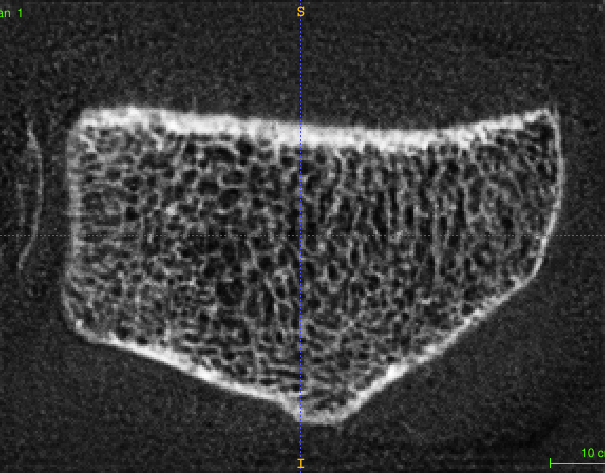}\hspace{.15cm}
		\includegraphics[height=.11\textheight]{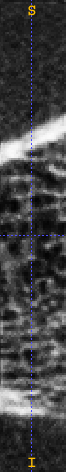}\hspace{-.05cm}
		\includegraphics[height=.11\textheight]{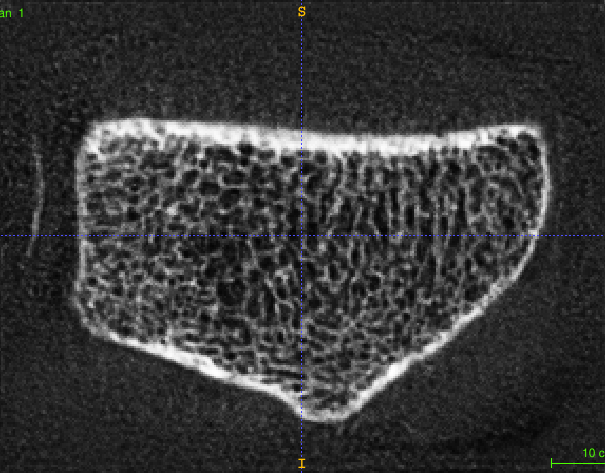}\hspace{.15cm}
		\includegraphics[height=.11\textheight]{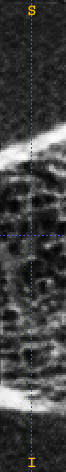}\hspace{-.05cm}
		\includegraphics[height=.11\textheight]{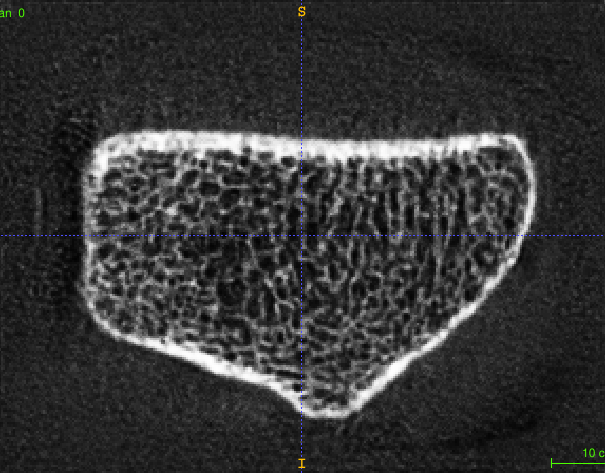}\hspace{.15cm}
		\includegraphics[height=.11\textheight]{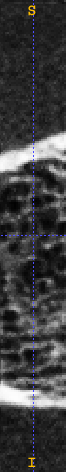}\hspace{-.05cm}
		\includegraphics[height=.11\textheight]{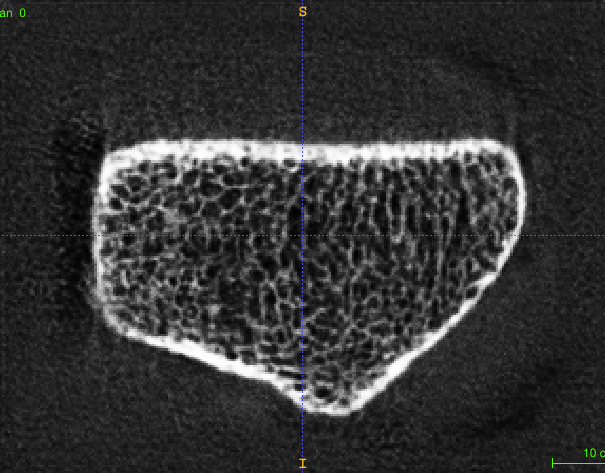}\vspace{-.06cm}
		
		\hspace{.25cm}
		\includegraphics[width=.197\textwidth]{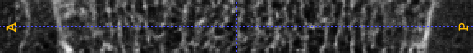}\hspace{.59cm}
		\includegraphics[width=.197\textwidth]{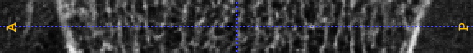}\hspace{.59cm}
		\includegraphics[width=.197\textwidth]{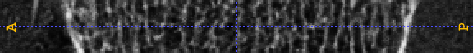}\hspace{.59cm}
		\includegraphics[width=.197\textwidth]{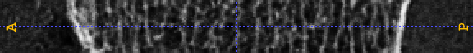}
		\vspace{.2cm}
		
		\caption{Synthetic HR-pQCT volumes sampled from the proposed 3D-StyleGAN approach with varying truncation levels. From left to right column: $\psi=\{1,0.7,0.4,0.1\}$.}
		\label{fig:stylegan_trunc}
	\end{figure}

	
	

\end{document}